\documentclass[conference]{IEEEtran}

\usepackage{comment}
\usepackage{soul}
\usepackage{spverbatim}
\usepackage{listings}
\usepackage{mdframed}
\usepackage{fancyvrb}
\usepackage{multirow}
\usepackage{minted}
\usepackage{multicol} 
\usepackage{tcolorbox}
\usepackage{colortbl}  
\usepackage{pifont}
\usepackage{makecell}
\usepackage{graphicx}
\usepackage{caption}
\usepackage{subcaption}
\usepackage{booktabs}
\usepackage{alltt}
\usepackage{rotating}
\usepackage{array}
\usepackage{multirow}
\usepackage{amsfonts}
\usepackage{amsmath}
\usepackage{xcolor}
\usepackage{siunitx}
\usepackage{algorithm}
\usepackage{algpseudocode}
\usepackage[normalem]{ulem}

\newcounter{remark}

\definecolor{navyblue}{rgb}{0.0, 0.0, 0.5}
\definecolor{darkgreen}{rgb}{0.0, 0.4, 0.0}

\newcommand{\sysname}{CCLab}

\newcommand{\para}[1]{{\vspace{3pt} \bf \noindent #1 \hspace{8pt}}}

\ifCLASSINFOpdf

\else

\fi

\begin{document}

\title{\sysname: Adversarial Testing of Learning- and Non-Learning-Based Congestion Controllers}

\author{\IEEEauthorblockN{Zhi Chen}
        \IEEEauthorblockA{UIUC\\
 		zhic4@illinois.edu}
        \and
        \IEEEauthorblockN{Shehab Sarar Ahmed}
        \IEEEauthorblockA{UIUC \& BUET\\
 		shehaba2@illinois.edu}
        \and
        \IEEEauthorblockN{Chenkai Wang}
        \IEEEauthorblockA{UIUC\\
 		chenkai3@illinois.edu}
        \and
        \IEEEauthorblockN{Brighten Godfrey}
        \IEEEauthorblockA{UIUC\\
 		pbg@illinois.edu}
        \and
        \IEEEauthorblockN{Gang Wang}
        \IEEEauthorblockA{UIUC\\
 		gangw@illinois.edu}}

\maketitle

\begin{abstract}

Congestion controllers (CCs) are critical to network performance, and yet their robustness under adverse conditions remains insufficiently understood. While recent learning-based CCs have demonstrated strong performance in controlled environments, it is unclear how they compare to traditional CCs when controllers' input signals are corrupted or when environmental conditions become systematically challenging.
In this paper, we introduce \sysname{}, an adversarial testing framework for systematically evaluating the robustness of both learning-based and non-learning-based CCs. \sysname{} includes a reinforcement learning (RL)-based adversarial agent that operates in a closed loop with the congestion control policy, generating bounded perturbations either on input signals (feature-level) or on external network conditions (environment-level), while preserving realism through explicit constraints. Using this framework, we compare learning-based CCs with non-learning-based CCs under both feature-level and environment-level adversarial conditions. While both types of CCs suffer from performance degradation under adversarial testing, we find that learning-based CCs, in general, are more robust than traditional human-designed algorithms. Finally, we show that our adversarial traces can be used to train more robust CCs that outperform existing learning-based CCs under both challenging and normal conditions.

\end{abstract}


\IEEEpeerreviewmaketitle

\section{Introduction}
\label{sec:intro}










Congestion controllers (CCs) play a central role in determining the performance and stability of modern networks. At the high level, CCs map the observed network signals (e.g., round-trip time, pacing rate, and estimated delay) to control actions (e.g., adjusting the size of the congestion window) to effectively utilize network resources while avoiding congestion collapse. Over multiple decades, CCs were manually designed by human experts using analytical models and heuristic control logic.
More recently, advances in machine learning have led to the emergence of {\em learning-based} congestion controllers~\cite{abbasloo2020classic,yang2024canopy,yen2023computers,jay2019deep,dong2018pcc} which have demonstrated promising performance in controlled environments.

While learning-based CCs often achieve high throughput and low latency under controlled benchmark testing, one key question remains before their wide deployment in real networks: {\em how robust are learning-based congestion controllers compared to traditional, non-learning-based controllers under adverse or unexpected conditions?} Prior evaluations~\cite{abbasloo2020classic,yang2024canopy,giacomoni2024reinforcement, mazilu2025learning} provide useful insights into typical behavior, but largely focus on performance under relatively benign conditions. As a result, it remains unclear how different CCs behave when network dynamics become misleading, highly challenging, or systematically unfavorable.

Answering this question is further complicated by the fundamental differences between learning-based and non-learning-based CCs in terms of their internal control logic and the input signals they rely on. 
Existing stress-testing approaches~\cite{giacomoni2024reinforcement, mazilu2025learning} often rely on predefined (manually created) evaluation scenarios where identical network conditions are applied to all CCs without adapting to the specific characteristics of each controller. This limits the ability to expose weaknesses that arise only under controller-specific worst-case conditions. 
Other works~\cite{ahmed26advnet,ray2022cc,robust-hotnet19} move beyond fixed scenarios by automatically generating network traces that degrade the performance of a given CC.
However, these works still focus on traditional, non-learning-based CCs without considering/comparing against {\em learning-based CCs}. Also, they all focus on modifying the network environmental conditions without evaluating the impact of the input signals to CCs.

\para{Our Approach.}
In this paper, we develop \sysname{}, an adversarial testing framework to systematically evaluate and compare the robustness of learning-based and non-learning-based congestion controllers. \sysname{} introduces an {\em adversarial agent} that operates in a closed loop with the congestion controllers. During the testing, a congestion controller continuously measures/observes network states (e.g., round-trip time, packet loss), and uses them as input signals to make congestion control decisions (e.g., by adjusting the congestion window). Meanwhile, the adversarial agent will also observe the network states and generate bounded perturbations that either (1) modify the direct input signals to the target congestion controller, or (2) change the external network conditions (which will indirectly influence the input signals used by the controller). The congestion controller itself remains unchanged and reacts solely to these modified observations. The adversarial objective is to construct challenging yet constrained conditions that systematically degrade performance while preserving plausibility and interpretability. 


Our adversarial testing consists of two types of manipulations. First, we perform {\em feature-level} manipulation, applying controlled modifications to a shared input signal to understand how different CCs respond to corrupted or noisy measurements. To enable consistent comparison, in this paper, we focus on an input signal that is widely used by both traditional and learning-based CCs: the minimum Round-Trip Time (RTT). More specifically, we test controller protocols that explicitly rely on RTT-related signals, including two learning-based CCs (Orca~\cite{abbasloo2020classic} and Canopy~\cite{yang2024canopy}) and representative delay-sensitive designs such as TCP Vegas~\cite{brakmo1994tcp} and BBR~\cite{cardwell2017bbr}. This feature-level testing provides an initial view of signal sensitivity and enables cross-policy comparison, but does not aim to model realistic attacker capabilities.



Second, we consider {\em environment-level} adversarial testing. Instead of directly manipulating the input signals (which can be unrealistic), we shape the external network conditions, such as bandwidth dynamics, which will {\em indirectly} influence the input signals to the congestion controllers. This also allows us to test and compare a broader set of congestion controllers, including two learning-based policies (Orca and Canopy) and four widely deployed non-learning-based controllers (Cubic~\cite{ha2008cubic}, TCP Vegas, TCP LP~\cite{kuzmanovic2003tcp}, and TCP Illinois~\cite{mazilu2025learning}). This covers key CC categories of loss-based, delay-based, and hybrid approaches. The adversarial traces are generated under explicit constraints (e.g., bounded smoothness), which serve as tunable factors that allow CC developers to explore a spectrum of challenging environments. 






\para{Evaluation and Findings.} Our analysis leads to a series of findings. We briefly highlight the following:    

{\em Feature-level perturbation} (\S\ref{sec:feature_level}). We observe that learning-based CCs have a higher level of robustness than non-learning-based CCs under feature-level perturbation. For example, under ±50\% adversarial noise on the input, Canopy degrades by only 3\% in bandwidth utilization, and Orca degrades by 8\%. In comparison, non-learning-based CCs have larger degradation, with BBR dropping by 15\% and TCP Vegas by 10\%. 

{\em Environment-level perturbation} (\S\ref{sec:env_level}). Under the more realistic environment-level perturbation, learning-based CCs continue to show higher robustness than non-learning-based CCs. For example, under adversarial bandwidth traces, the degradation of learning-based CCs is modest (12\% for Orca and 7\% for Canopy) but it is more severe for non-learning-based CCs (19\% for Cubic, 17\% for TCP Vegas, 18\% for TCP Illinois, and 30\% for TCP LP). 


{\em Learning-based vs. non-learning-based CCs} (\S\ref{sec:env_reason}, \S\ref{sec:case_study}).
We further analyze the potential reasons behind the different robustness levels of learning and non-learning-based CCs. One hypothesis is that the improved robustness of learning-based CCs may stem from their {\em smoother and more continuous action space}. However, our analysis shows the {\em opposite} direction, as learning-based CCs make more drastic congestion window (cwnd) adjustments than non-learning-based CCs. We further analyze the {\em responsiveness} of cwnd adjustments of these protocols. We show that learning-based CCs are more responsive to bandwidth changes with more precise cwnd adjustments, while non-learning-based CCs rely on rigid control rules that limit their responsiveness. This offers an explanation for their robustness under adversarial bandwidth conditions. Finally, we perform a case study on a non-learning-based CC (TCP LP) that has the most performance degradation. We illustrate how adversarial bandwidth variations consistently trigger TCP LP's conservative design (e.g., early congestion avoidance) to prevent it from using available bandwidth. 

{\em Adversarial training} (\S\ref{sec:adv_train}). Finally, we show that the adversarial traces generated by \sysname{} can effectively improve the robustness of a learning-based CC. For example, adversarial training increases Orca's bandwidth utilization under adversarial conditions by 6\% (from 85.49\% to 91.66\%) without increasing queuing delay. Also, it maintains a comparable performance on benign workloads (with less than 1\% changes). This illustrates the value of \sysname{} to CC developers for improving their protocols.

\para{Contributions.}
We make the following contributions:
\begin{itemize}
    \item {\em A New Evaluation Framework.}  
    We introduce \sysname{}, an evaluation framework to compare the robustness of learning and non-learning based CCs. Our experiments show that learning-based CCs exhibit stronger robustness than traditional CCs.
    
     \item {\em A New Network Trace Generation Method.}  
    We develop an adversarial agent that can generate challenging network conditions while allowing developers to control key characteristics of the adversarial traces (e.g., the level of dynamics, targeted metrics). This enables controlled and reproducible stress testing against CCs.
    
    \item {\em A New Tool for CC Developers.}
    We demonstrate how \sysname{} can help developers identify failure cases, understand performance degradation, and improve CC policies through techniques such as adversarial retraining. To support future research and practical use, we will release our code with this paper.
    
\end{itemize}

\section{Background and Related Work}
\label{sec:related}

Congestion controllers (CCs) are fundamental components of modern network transport protocols, responsible for regulating the sending rate of data flows to efficiently utilize network resources while avoiding congestion collapse. At their core, CCs dynamically adjust control variables such as the congestion window ($cwnd$) or pacing rate based on observed network feedback, such as acknowledgments, delay measurements, and packet loss signals. Over the past decades, a wide range of CCs have been developed, broadly spanning traditional human-designed protocols based on analytical models and heuristics, as well as more recent learning-based approaches that leverage data-driven decision-making.



\subsection{Non-learning-based Congestion Controllers}
\label{sec:related-nonlearning-based}
Traditional CCs are typically manually designed, based on analytical models and heuristic rules. 


\para{Loss-based Algorithms.}
Loss-based CCs interpret {\em packet loss} as the primary indicator of congestion. They introduce different variants of the additive-increase multiplicative-decrease (AIMD) paradigm. TCP Reno represents the canonical design. Improvements over Reno include high-speed adaptations such as BIC~\cite{xu2004binary}, Cubic~\cite{ha2008cubic}, HighSpeed TCP~\cite{floyd2003highspeed}, H-TCP~\cite{leith2004h}, Hybla~\cite{caini2004tcp}, and Scalable TCP~\cite{kelly2003scalable}, which modify the window growth function to better utilize high bandwidth-delay product (BDP) networks. These algorithms are widely deployed due to their simplicity and robustness, but their reliance on packet loss as a congestion signal often leads to persistent queue buildup and increased latency.

\para{Delay-based Algorithms.}
Delay-based CCs infer congestion from variations in round-trip time (RTT), aiming to detect queue buildup before packet loss occurs. TCP Vegas~\cite{brakmo1994tcp} is a representative design that adjusts its sending rate based on deviations between expected and observed throughput. Subsequent variants such as TCP Veno~\cite{fu2003tcp} and CDG~\cite{hayes2011revisiting} extend this idea by incorporating additional heuristics to improve robustness under noisy conditions. While delay-based approaches can achieve low latency and stable behavior, they are sensitive to measurement noise and may under-utilize available bandwidth.

\para{Hybrid Algorithms.}
Hybrid CCs combine multiple congestion signals, typically integrating both loss and delay information to balance utilization and latency. TCP Illinois~\cite{liu2006tcp} and TCP YeAH~\cite{baiocchi2007yeah} adapt their window adjustment rules based on both delay trends and packet loss events, while TCP Westwood~\cite{mascolo2001tcp} incorporates bandwidth estimation to refine its response after congestion. TCP LP~\cite{kuzmanovic2003tcp} further extends this design space by leveraging one-way delay measurements to detect early congestion and yield bandwidth to competing flows. These hybrid approaches aim to improve adaptability across diverse network conditions, but their performance depends on the interaction between multiple control signals and parameters.

\para{Explicit-state Estimation Designs.}
Instead of directly reacting to the observed signals, some CCs estimate underlying network state variables, such as bottleneck bandwidth, propagation delay, or queue occupancy, and derive control actions from these estimates. For example, TCP BBR~\cite{cardwell2017bbr} continuously estimates the bottleneck bandwidth and minimum round-trip time (RTT), and sets its sending rate based on the inferred bandwidth-delay product (BDP), rather than directly responding to packet loss or delay increase. Similarly, DCTCP~\cite{alizadeh2010data} uses explicit congestion notification (ECN) signals to approximate the level of queue occupancy and adjusts its congestion window proportionally to the extent of congestion.

\subsection{Learning-based Congestion Controllers}
Recent advances in machine learning have led to a new class of congestion controllers that learn control policies directly from data rather than relying on handcrafted rules. These approaches typically formulate congestion control as a sequential decision-making problem, where a policy maps observed network statistics to control actions. 

Early work has explored the use of reinforcement learning (RL) to learn congestion control policies. For example, Aurora~\cite{jay2019deep} formulates congestion control as an RL problem, where a neural policy maps a history of network observations (e.g., latency, throughput, and loss) to rate adjustments. PCC Vivace~\cite{dong2018pcc} adopts an online learning approach that continuously evaluates the utility of actions and updates the sending rate via gradient-based optimization. Glider~\cite{xia2023glider} applies deep reinforcement learning to directly learn congestion window adjustment policies by discretizing continuous network interactions into decision steps.

Recently, hybrid designs combine machine learning with rule-based control, where learning may support state estimation or directly influence control decisions. MLACC~\cite{elbery2023toward} uses a supervised model to estimate bottleneck utilization while relying on heuristic control for rate adjustment. Orca~\cite{abbasloo2020classic} integrates a deep RL agent into a TCP framework to periodically adjust the congestion window. Building on this line of work, Canopy~\cite{yang2024canopy} addresses the sensitivity and instability of learning-based policies under unforeseen inputs by incorporating {\em formal verification} into the learning loop, providing robustness and performance guarantees.

Finally, researchers have also explored learning from existing congestion control algorithms. Sage~\cite{yen2023computers} proposes an offline reinforcement learning framework that learns a new policy from datasets collected from diverse non-learning-based CCs. 


\subsection{Benchmarks for Congestion Controllers}
Researchers have examined the performance of CCs but prior works have largely focused on {\em average-case} behavior under fixed or benign conditions, leaving {\em worst-case} robustness insufficiently explored.

Giacomoni et al.~\cite{giacomoni2024reinforcement} present one of the first large-scale evaluations of reinforcement learning (RL)-based congestion controllers. Their results show that RL-based approaches can quickly utilize available bandwidth but often suffer from unfairness and degraded performance under dynamic conditions. Mazilu et al.~\cite{mazilu2025learning} extend this line of work by conducting an in-depth comparison between learning-based and human-designed CCs, highlighting generalization and instability issues when operating outside training regimes. However, both studies rely on predefined scenarios, fixed traces, or manually constructed benchmarks to evaluate controller behavior: they do not systematically explore the space of challenging network conditions or worst-case scenarios. 

To move beyond manually constructed scenarios, researchers propose to automatically generate network traces that degrade CC performance~\cite{ahmed26advnet,ray2022cc,robust-hotnet19}. By evolving traces using performance-driven fitness functions, they adapt the generated scenarios to the specific congestion controller to discover unexpected (and often non-obvious) behaviors. 
Compared with existing work, our paper has {\em three novel aspects}. First, existing works~\cite{ahmed26advnet,ray2022cc,robust-hotnet19} still focus on traditional CCs. They did not focus on comparing learning-based CCS with non-learning-based CCs. In addition, they mainly focus on modifying network environment conditions without evaluating the impact of input signals of CCs (i.e., feature-level perturbations). Finally, they did not use adversarial training to improve learning-based CCs.

\section{Adversarial Testing Framework}
\label{sec:method}

\begin{figure}[t]
\centering
\begin{minipage}[t]{0.5\textwidth}
\includegraphics[width=0.99\textwidth]{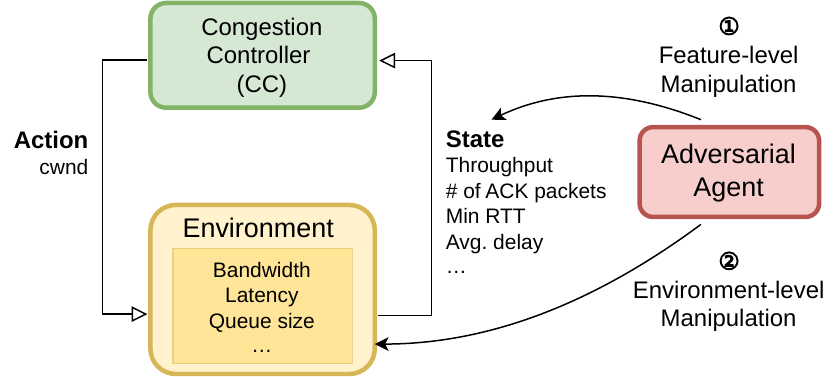}
\end{minipage}
\caption{
\sysname: our adversarial testing framework for the congestion controller (CC). The CC can be learning-based or non-learning-based. ``cwnd'' means ``congestion window.''   
}
    \label{fig:framework}
    \vspace{-0.12in}
\end{figure}



In this section, we provide an overview for \sysname. We use reinforcement learning (RL) as the adversarial engine to construct our testing framework for a few considerations. First, congestion controllers (CCs) naturally formulate an RL-like interaction loop with the network, where network signals influence CC's actions, and the actions will further influence the network conditions in the future. An ``adversarial'' RL agent can be easily plugged into this loop to capture and learn from such dynamic interactions. Second, using RL allows us to explicitly encode adversarial objectives and constraints through reward design and bounded action spaces. This allows us to create ``targeted'' adversarial traces to stress test CCs to expose failure modes and perform diagnosis. Ultimately, we want to provide testing tools for CC protocol designers to better understand the weaknesses of their designs.  

\para{Framework Overview.}
Fig.~\ref{fig:framework} illustrates the high-level architecture of our testing framework, \sysname{}. 

Congestion control is formulated as a sequential decision-making problem. A congestion controller (CC) in the green box will collect network states (e.g., min RTT) from the kernel at a fixed monitoring interval. The network states are passed to the user-space, where the CC produces control actions (e.g., adjustments to the cwnd). The actions are then enforced at the transport stack while the underlying TCP logic continues its fine-grained ACK-driven updates. Here, the CC can be either a learning-based protocol or a non-learning-based protocol. 


As shown in Fig.~\ref{fig:framework}, we introduce an adversarial agent (red box) into this observation-control pipeline. The adversarial agent seeks to manipulate the states observed by the target CC to influence its actions. The congestion controller itself remains unchanged and reacts solely to the perturbed input network states (i.e., input signals) it receives. In this paper, we consider two ways to manipulate the input network states: (1) feature-level manipulation, which directly perturbs the input signals used by the CC (e.g., RTT measurements), and (2) environment-level manipulation, which modifies underlying network conditions (e.g., bandwidth) to indirectly influence these observations. From a systems perspective, the adversarial agent operates externally to the congestion controller, without modifying protocol internals such as congestion window updates, pacing mechanisms, or state transitions. This design preserves the original behavior of each congestion controller while enabling systematic and controlled adversarial testing.

\para{Threat Model.}

\sysname{} explicitly supports two complementary control surfaces as shown in Fig.~\ref{fig:framework}. In (1), the adversarial agent performs \emph{feature-level manipulation} by directly perturbing input signals (e.g., min RTT) before they are consumed by the congestion controllers. This provides a controlled diagnostic setting to probe how sensitive different CCs are to corrupted or noisy measurements. In (2), the agent performs \emph{environment-level manipulation} by modifying environment variables, such as bandwidth or link delay, thereby indirectly influencing the observations for the CCs.


These two types of adversarial testing serve complementary purposes. Feature-level manipulation isolates the dependence of CCs on the specific input signals and enables cross-policy comparison. In contrast, environment-level manipulation reflects more {\em realistic conditions}, where performance degradation arises from adverse network dynamics rather than direct input corruption. We will provide more detailed justifications for these two adversarial testing types  (Sections~\ref{sec:feature_level} and \ref{sec:env_level}).

\para{Agent Training.}
The adversarial agent is trained online using reinforcement learning. At each monitoring interval, it observes network states and produces actions that modify a designated control surface, with the resulting system behavior determining the adversarial RL reward.
To regulate adversarial strength, all actions are subject to explicit constraints that bound their magnitude, temporal smoothness, or long-term variability. These constraints act as tunable parameters, enabling systematic exploration of different adversarial regimes. By adjusting them, CC developers can examine how policies respond under varying degrees of challenge, from mild perturbations to highly structured adversarial conditions. Specific control surfaces and constraint configurations are described in the corresponding experimental sections.



\para{Implementation and Evaluation Metrics.}
Instead of using network simulators (e.g., NS3), we implement the framework on a network emulator Mahimahi~\cite{netravali2015mahimahi} to experiment with more realistic network environments and test real CC implementations. This emulator is used by many existing learning-based CCs~\cite{abbasloo2020classic, yang2024canopy,ahmed2026advnet} as it supports real packet transmissions over emulated links, enabling faithful evaluation of congestion control behavior. We use the network traces from~\cite{yang2024canopy}, which capture a wide range of dynamic network conditions. 
Following the standard setup in prior work~\cite{yang2024canopy}, each trace is emulated using Mahimahi with a fixed link delay of 10\,ms and a buffer size set to $2\times$BDP, representing a typical operating regime. Unless otherwise specified, this configuration is used consistently across all experiments.

We measure performance using two primary metrics: {\em bandwidth utilization} and {\em queuing delay}. Bandwidth utilization captures how effectively a CC utilizes available bandwidth (higher is better), while queuing delay reflects its responsiveness and stability under congestion (lower is better). Together, these metrics characterize the key trade-offs in congestion control performance. All results are reported after adversarial training has converged and are averaged over three runs to account for stochasticity in both network dynamics and adversarial learning.

\section{Feature-level Adversarial Testing}
\label{sec:feature_level}

We begin our evaluation with a feature-level sensitivity analysis. The goal of this stage is not necessarily about modeling realistic adversarial capabilities. Instead, we seek to obtain an initial understanding of how congestion control algorithms respond to corrupted, biased, or extreme input signals. It allows us to directly probe the dependence of different policies on specific signals. From a protocol design perspective, this approach can serve as a stress-testing tool by injecting arbitrary feature values, enabling designers to systematically evaluate robustness under extreme or unexpected input conditions. We acknowledge that this setting may not be ideal to model attackers (e.g., it makes strong assumptions on attacker capabilities, such as compromising the measurement components or directly interfering with internal signal processing). Instead, we use this setting to capture undesired scenarios such as measurement noise, estimation bias, or transient inaccuracies in signal computation. The analysis serves as a diagnostic tool to measure feature sensitivity and identify signals that are critical to policy behavior. 






\subsection{Feature-level Control Surface}
As discussed in Section~\ref{sec:related}, learning and non-learning-based CCs rely on different sets of input signals (or features). This is especially true for traditional, human-designed CCs, as they often rely on one or a few distinct input signals to make decisions (e.g., loss-based algorithms mainly monitor package loss while completely ignoring delay features such as RTT). As such, it is difficult to manipulate certain features while fairly comparing different algorithms. To enable cross-policy comparison, our idea is to identify {\em common input signals} explicitly or implicitly used by most learning and non-learning based CCs and explore how they react to the manipulations on these features. In this paper, we focus on minimum RTT (or min-RTT) as it plays an important role in delay estimation, pacing decisions, and congestion window adjustment, and is widely used in delay-based and hybrid congestion control algorithms. Also, it is a key input signal considered by all learning-based CCs. We understand that min-RTT does not universally apply to all CCs, particularly those that only rely on loss-based signals. In this paper, we use min-RTT to demonstrate our adversarial agent design. The same methodology can be directly applied to other input signals to design adversarial agents (pacing rate, acknowledgment statistics) to measure their influence on CCs that depend on them. This flexibility allows protocol developers to adapt the framework to different subsets of congestion control algorithms based on their specific design characteristics.

\subsection{Adversarial Agent Design}
\label{sec:feature-design}

CCs are optimizing their decisions to achieve high bandwidth utilization while reducing queuing delay (and avoiding network congestion). Most learning-based CCs design their reward function to achieve these goals. For instance, Orca and Canopy use the following reward function to jointly capture throughput efficiency, packet loss penalty, and delay sensitivity. The reward at time step $t$ is defined as:
\begin{equation}
R_t = \frac{T_t - \lambda L_t}{B_{\max}} \cdot D_t
\end{equation}
where $T_t$ denotes the achieved throughput, $L_t$ is the packet loss rate, $\lambda$ is the loss penalty coefficient, and $B_{\max}$ represents the maximum available bandwidth used for normalization.

The delay-related factor $D_t$ is defined as:
\begin{equation}
D_t =
\begin{cases}
\frac{\gamma \cdot RTT_{\min}}{RTT_t}, & \text{if } \gamma \cdot RTT_{\min} < RTT_t \\
1, & \text{otherwise}
\end{cases}
\end{equation}
where $RTT_t$ is the smoothed round-trip time, $RTT_{\min}$ is the minimum observed RTT, and $\gamma$ is a delay margin coefficient.

\para{An Naive Adversarial Agent Design.}
The adversarial objective is to degrade the utilization--delay trade-off achieved by the congestion controller. With this consideration, a {\em naive} design is to simply define the adversarial reward as {\em the negative} of the original reward used to train learning-based controllers: 
\begin{equation}
R^{\text{adv}}_t = - R_t
\end{equation}
thereby encouraging behaviors that move the system away from its intended operating point.


In terms of capability, we model the adversary as performing bounded perturbations on input signals to reflect uncertainty in network measurements.
Specifically, we perturb the estimated minimum RTT within a constrained range around its true value. We consider both small perturbations (e.g., around 5\%), which capture realistic measurement noise and estimation bias, and larger perturbations (up to 50\%), which are used to stress-test controller behavior under stronger adversarial influence. The latter does not aim to model typical noise levels, but rather to explore the limits of policy robustness when the adversarial capability is significantly amplified. 
Formally, the perturbed minimum RTT lies within $\text{minRTT} \cdot [1 - x\%, 1 + x\%]$, where $x$ controls the perturbation magnitude.

\begin{figure}[t]
\centering
\begin{minipage}[t]{0.4\textwidth}
\includegraphics[width=0.99\textwidth]{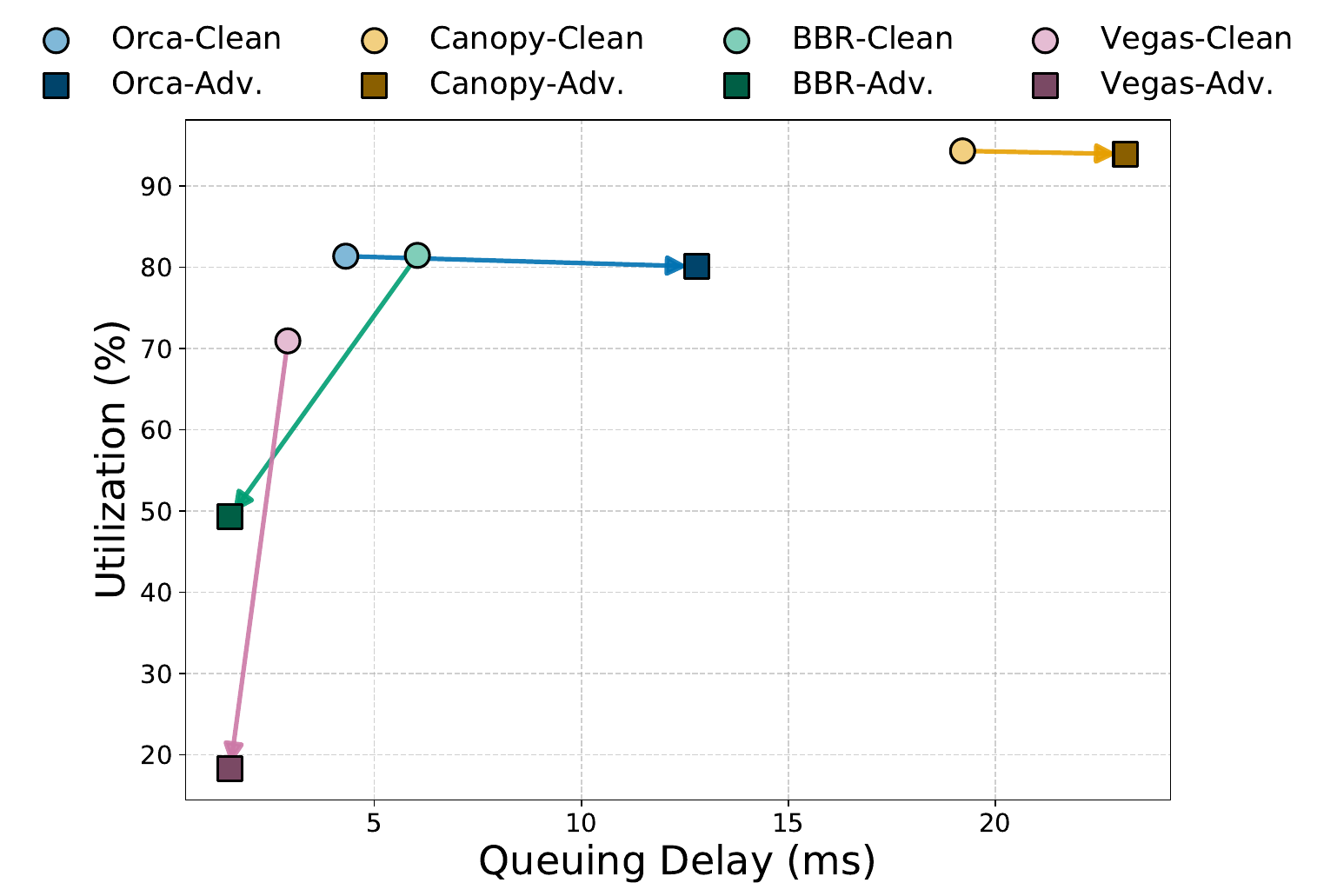}
\end{minipage}
\caption{
Feature-level adversarial testing results with 50\% adversarial capacity, under the naive reward design. 
}
\label{fig:feature05}
\vspace{-0.12in}
\end{figure}

We implemented this adversarial agent and performed preliminary experiments. The results highlight a {\em key limitation} of this naive approach. More specifically, the adversarial agent tries to simultaneously attack bandwidth utilization and queuing delay (e.g., reducing utilization and increasing delay). At the same time, bandwidth utilization and queuing delay have an inherent trade-off between themselves (e.g., lower bandwidth utilization often reduces queuing delay). As a result, the agent may attack different CCs along different dimensions, making it difficult to compare and interpret the degradation across CCs. Figure~\ref{fig:feature05} illustrates this problem. We show that the agent mainly degrades the performance of Canopy and Orca by increasing their queuing delay. However, for BBR and Vegas, the agent managed to significantly degrade their bandwidth utilization, but with reduced queuing delay as well. Due to the inherent trade-off between bandwidth utilization and queuing delay, it is unclear whether the degradation of bandwidth utilization reflects a true performance degradation from the adversarial agent or a natural trade-off from the reduced queuing delay. It is also difficult to compare the performance of BBR and Vegas with the other protocols.

\para{The Improved Design of Adversarial Agent.}
To address this problem, we want to have more explicit controls on the adversarial agent's behaviors. 
Instead of jointly optimizing multiple competing metrics, we primarily focus on utilization as the optimization objective, while treating queuing delay as a constraint. Specifically, we require that the queuing delay under adversarial conditions {\em should be no lower than} the original/baseline performance while the agent optimizes to degrade the bandwidth utilization. This is to ensure that any degradation in bandwidth utilization is not achieved by trivially reducing congestion.

Here, we take {\em bandwidth utilization} as the main attack target and make {\em queuing delay} a constraint. This mainly considers their practical impact on service quality in real-world scenarios. In other words, bandwidth is a more important factor and a more realistic adversary's attack goal. Today, most CCs for TCP are optimized for bulk reliable data transfer applications in fast, long-distance networks where bandwidth utilization is a higher-prioritized goal. In addition, RTT usually overwhelms queuing delay by magnitude, and most TCP applications are not latency-sensitive. As such, taking queuing delay as the primary attack target is not well justified.

To reflect this idea in a new reward function, let $U_t$ denote the utilization at time step $t$. We define the queuing delay as the difference between the smoothed RTT and the minimum RTT:
\begin{equation}
d_t = RTT_t - RTT_{\min}.
\end{equation}

To improve the robustness against transient fluctuations while retaining responsiveness to instantaneous changes, we maintain a short history of recent queuing delays over a sliding window of size $H$. Based on this history, we compute a local average delay and a near-instantaneous delay estimate:
\begin{equation}
\bar{d}_t = \frac{1}{H} \sum_{i=t-H+1}^{t} d_i, \quad
\tilde{d}_t = \frac{1}{K} \sum_{i=t-K+1}^{t} d_i,
\end{equation}
where both $H$ and $K$ are small constants (e.g., $H=5$, $K=1$). Here, $\bar{d}_t$ captures short-term averaged behavior over the recent sliding window, while $\tilde{d}_t$ reflects the most recent observed delay. This design avoids sensitivity to noise while still enabling the reward to react quickly to current network conditions.

We then define a delay-based penalty term $R^{\text{delay}}_t$ using a threshold $\tau$:
\begin{equation}
R^{\text{delay}}_t =
\begin{cases}
-\alpha, & \text{if } \bar{d}_t < \tau \ \text{and} \ \tilde{d}_t < \tau, \\
0, & \text{otherwise},
\end{cases}
\end{equation}
where $\alpha > 0$ is a constant penalty coefficient. The threshold $\tau$ is derived from a baseline performance. Here, the baseline refers to the original performance of the algorithm without manipulation. 
The average queuing delay under baseline is $\tau$, which reflects the CC's performance in a non-adversarial setting. This ensures that the adversarial agent is required to reach at least this baseline level of delay before optimizing for utilization degradation.

Finally, the overall adversarial reward is defined as:
\begin{equation}
R^{\text{env}}_t = -U_t + R^{\text{delay}}_t
\end{equation}

This design encourages the adversarial agent to operate in a regime where queuing delay has already reached a predefined threshold, and then primarily optimizes for degradation in utilization. The lower-bound condition ($\bar{d}_t < \tau$ and $\tilde{d}_t < \tau$) prevents the agent from exploiting low-delay scenarios, where reduced utilization could be achieved without inducing meaningful congestion. By enforcing this constraint, we effectively decouple the impact of the utilization-delay trade-off, and the
resulting changes in utilization can be interpreted unambiguously as performance degradation.

\subsection{Evaluation Setup}


We apply the testing framework against a representative set of congestion controllers (CCs), including two learning-based policies (Orca and Canopy) and two non-learning-based designs (BBR and TCP Vegas). These protocols are selected because they all rely on minimum RTT estimation as part of their control logic. This shared dependency enables meaningful feature-level perturbation and allows us to directly compare their sensitivity to this input signal.
For each CC, we first use reinforcement learning to train an adversarial agent in an online manner, where the agent interacts with the running protocol, dynamically manipulating the input feature (i.e., minimum RTT) at each RL step. After training, we then run the trained adversarial agent against the CC during runtime. 



We evaluate the impact of feature-level perturbations under two adversarial capability levels, corresponding to $\pm5\%$ and $\pm50\%$ perturbation ranges on the minimum RTT signal. Each setting includes two main conditions: \textit{Clean} (without attack), and \textit{Adversarial} (with attack). Additionally, we run a \textit{Noise} condition (i.e., random perturbation within the same range). This noise setting (with random perturbation), unfortunately, could not control the utility-delay trade-off, and thus the results are only presented in Appendix~\ref{sec:feature_level_noise} for reference.

All evaluations are conducted with the network trace dataset used by Canopy~\cite{yang2024canopy}. 
The dataset contains 18 simulated traces and 3 real-world traces. The simulated traces are bandwidth traces used to emulate time-varying network conditions in Mahimahi, enabling controlled yet diverse evaluation of congestion control behavior under dynamically changing link capacities. The real-world traces are derived from cellular network measurements (e.g., commercial LTE networks), capturing realistic bandwidth variability observed in operational environments. 
In this section, we mainly report our results on the simulated traces, as these traces cover more diverse and challenging scenarios. We also performed additional experiments on real-world traces in Appendix~\ref{sec:feature_level_real} --- the conclusions remain consistent. Here, we randomly select 4 traces for training the agent, and perform testing on the remaining 14 traces. To ensure stability and reproducibility, each experiment is repeated three times per trace, and we report performance metrics averaged across runs and traces.

Unless otherwise specified, we fix the remaining environment parameters, following the same configurations used in previous work~\cite{yang2024canopy}. In particular, the base link delay is set to $10\,\mathrm{ms}$, and the queue size is configured as twice the bandwidth-delay product (2 BDP).

\begin{table}[t]
    \centering
    \begin{tabular}{llccc}
        \toprule
        \textbf{Model} & \textbf{Settings} & \textbf{Util. ($\downarrow$)} & \textbf{Delay ($\uparrow$)} & \textbf{P95 delay ($\uparrow$)} \\
        \midrule
             & Clean       & 81.36 & 4.31  & 12.51 \\
        Orca 
             & Adversarial & 80.48 & 4.50 & 19.00 \\
        \midrule
             & Clean       & 94.32 & 19.21 & 45.53 \\
        Canopy   
             & Adversarial  & 94.22 & 20.30 & 46.61 \\
        \midrule
            & Clean & 81.46 & 6.04 & 22.17 \\
        BBR     
             & Adversarial & 76.94 & 6.65 & 26.04 \\
        \midrule
             & Clean & 70.94 & 2.91 & 6.45 \\
        Vegas   
             & Adversarial & 70.82 & 2.97 & 6.50  \\
        \bottomrule
    \end{tabular}
    \caption{Feature-space manipulation results under minRTT perturbation ($[0.95, 1.05]$). 
    }
    \label{tab:feature_minrtt_005_newreward}
\end{table}

\begin{table}[t]
    \centering
    \begin{tabular}{llccc}
        \toprule
        \textbf{Model} & \textbf{Settings} & \textbf{Util. ($\downarrow$)} & \textbf{Delay ($\uparrow$)} & \textbf{P95 delay ($\uparrow$)} \\
        \midrule
        Orca     & Clean     & 81.36 & 4.31  & 12.51 \\
             & Adversarial & 73.67 & 4.23 & 16.80  \\
        \midrule
        Canopy     & Clean     & 94.32 & 19.21 & 45.53 \\
             & Adversarial & 91.70 & 22.22 & 49.47 \\
        \midrule
        BBR     & Clean & 81.46 & 6.04 & 22.17 \\
             & Adversarial  & 66.78 & 6.25 & 26.78 \\
        \midrule
        Vegas     & Clean & 70.94 & 2.91 & 6.45 \\
             & Adversarial & 60.33 & 2.74 & 4.42  \\
        \bottomrule
    \end{tabular}
    \caption{Feature-space manipulation results under minRTT perturbation ($[0.5, 1.5]$). }
    \label{tab:feature_minrtt_05_newreward}
\end{table}

\begin{figure}[t]
\centering
\begin{minipage}[t]{0.4\textwidth}
\includegraphics[width=0.99\textwidth]{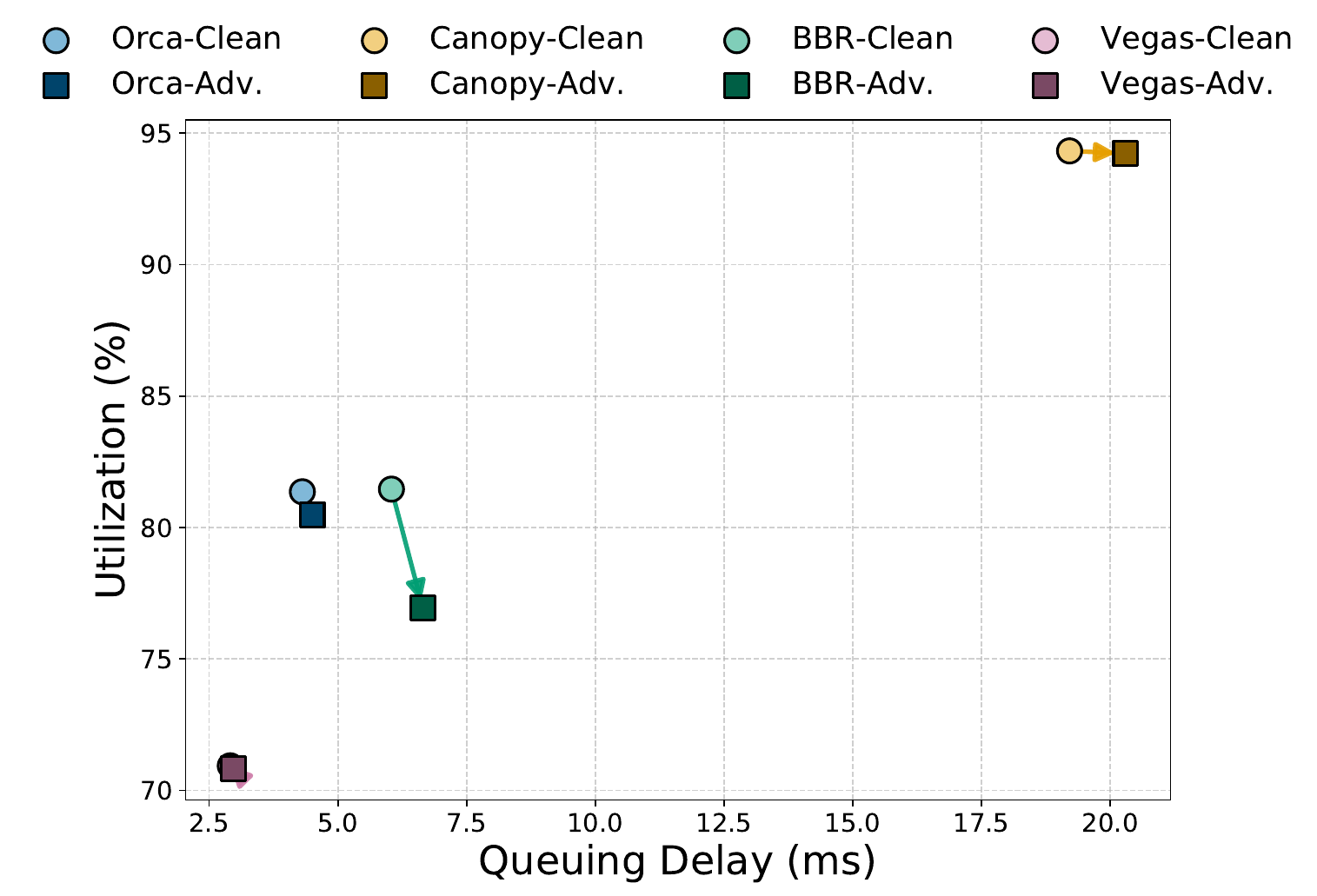}
\end{minipage}
\caption{
Feature-level adversarial testing results with 5\% adversarial capacity (new reward applied). 
}
    \label{fig:feature005_newreward}
    \vspace{-0.12in}
\end{figure}

\begin{figure}[t]
\centering
\begin{minipage}[t]{0.4\textwidth}
\includegraphics[width=0.99\textwidth]{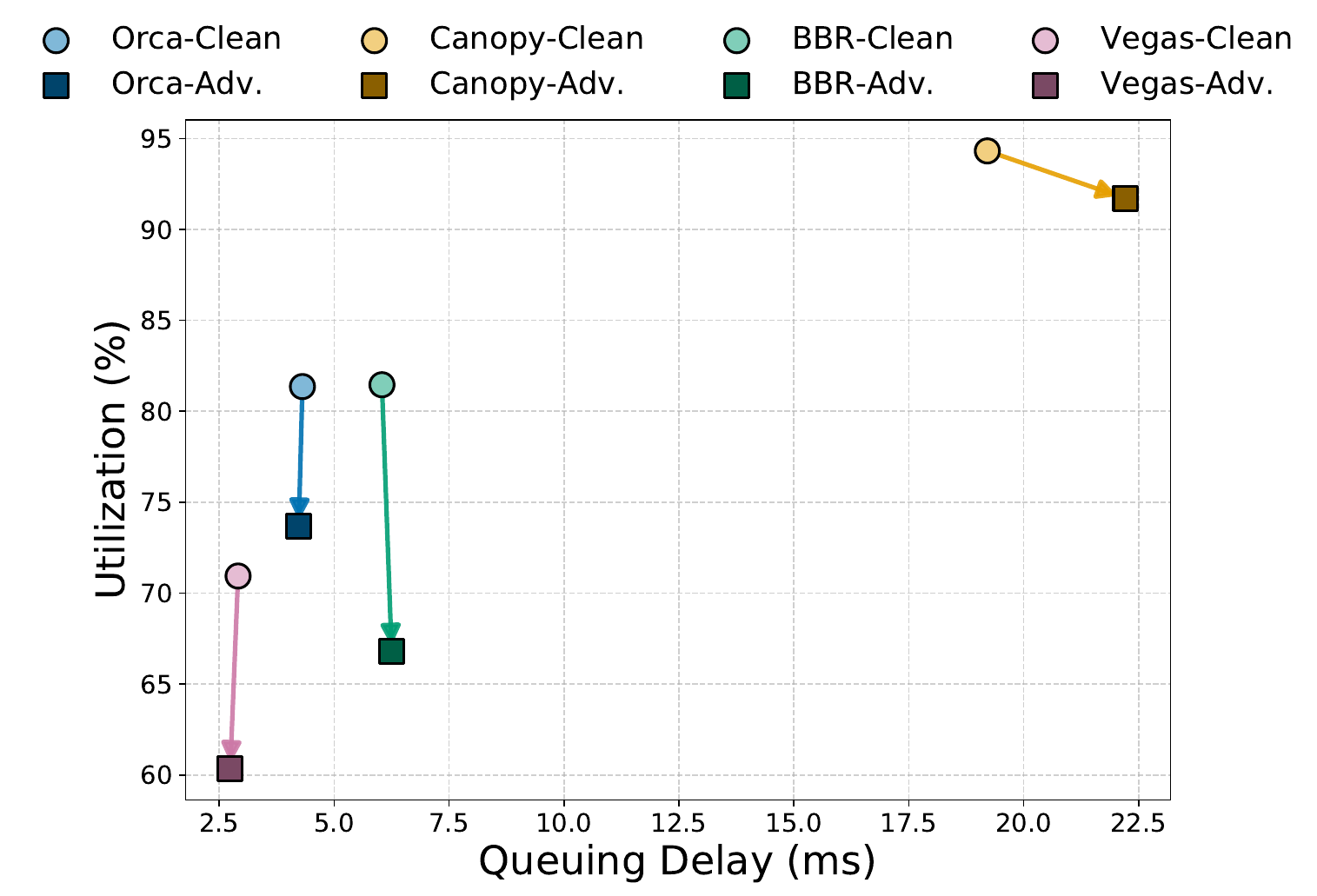}
\end{minipage}
\caption{
Feature-level adversarial testing results with 50\% adversarial capacity (new reward applied).
}
    \label{fig:feature05_newreward}
    \vspace{-0.12in}
\end{figure}

\subsection{Performance and Findings}

The results are presented in Tables~\ref{tab:feature_minrtt_005_newreward} and~\ref{tab:feature_minrtt_05_newreward}. 
We start by verifying whether the queuing delay constraint is consistently enforced. As shown in Tables~\ref{tab:feature_minrtt_005_newreward} and~\ref{tab:feature_minrtt_05_newreward}, the queuing delay under adversarial settings is no lower than (or comparable to) the clean baseline across both perturbation ranges. This confirms that the reward function helps to eliminate the ambiguity introduced by the utilization–delay trade-off, allowing us to isolate utilization degradation as the primary attack target. 

Under small perturbations (±5\%), the overall trends remain consistent with our earlier observations. 
Figure~\ref{fig:feature005_newreward} shows that Orca, Canopy, and TCP Vegas maintain utilization levels very close to their clean baselines, indicating limited sensitivity to mild feature corruption. In contrast, BBR exhibits a modest degradation of around 5\%, suggesting a higher sensitivity even under relatively small perturbations.

As the adversarial capability increases to ±50\%, the differences across policies become more pronounced. From Table~\ref{tab:feature_minrtt_05_newreward}, Orca experiences an approximately 8\% drop in utilization, while Canopy shows a smaller degradation of around 3\%. In comparison, non-learning-based policies are more significantly affected: BBR suffers a substantial reduction of about 15\%, and TCP Vegas shows a degradation of roughly 10\%. These trends are further illustrated in Figure~\ref{fig:feature05_newreward}, where the separation between learning-based and non-learning-based policies becomes clearly visible when delay is controlled. 

Overall, these results highlight that different CCs rely on shared input signals in fundamentally different ways, leading to heterogeneous sensitivity under adversarial perturbations. In particular, under feature-level manipulation of the minimum RTT signal, the learning-based policies (Orca and Canopy) demonstrate stronger robustness than the selected non-learning-based designs (BBR and TCP Vegas), primarily by maintaining stable utilization even under large perturbations. More broadly, the feature-level adversarial testing framework provides a controlled and systematic means to isolate and stress specific input dependencies, enabling developers to pinpoint vulnerabilities and better understand the robustness characteristics of their designs.

\begin{tcolorbox}[width=\linewidth, colback=white!95!black, boxrule=0.5pt, left=2pt,right=2pt,top=1pt,bottom=1pt]
\stepcounter{remark}
{\bf Remark \arabic{remark}:}
{
Under feature-level manipulation, most CCs remain robust to small perturbations (5\% of adversarial noise). As we increase the perturbation magnitude, learning-based CCs exhibit a higher level of robustness than non-learning-based CCs. 
}
\end{tcolorbox}

\section{Environment-level Adversarial Testing}
\label{sec:env_level}

While feature-level sensitivity analysis provides useful diagnostic insights, it relies on direct manipulation of policy inputs and therefore may not reflect realistic deployment scenarios. In practice, congestion control policies are rarely exposed to adversarially crafted feature values. Instead, performance degradation typically arises from adverse or rapidly changing network conditions, such as fluctuating available bandwidth, competing traffic, or changes in end-to-end delay caused by routing or queuing dynamics.

To evaluate robustness under more realistic conditions, we conduct environment-level adversarial testing. In this stage, the adversarial agent does not manipulate policy inputs directly. Instead, it influences external network conditions that indirectly shape the observations available to the congestion control policy. This setting preserves the integrity of the policy’s internal logic and measurement pipeline, while allowing the adversary to induce challenging operating conditions through plausible network dynamics.

\subsection{Environment-level Control Surfaces}
In environment-level experiments, the adversarial agent operates on network parameters that are external to the congestion control algorithm but fundamentally define the environment in which the policy operates. We consider two such parameters: available bandwidth and link delay. These variables are not directly controlled by the policy, yet they shape the end-to-end measurements (e.g., throughput and RTT) that drive congestion control decisions.

Bandwidth variation captures a wide range of real-world phenomena, including cross traffic, traffic engineering decisions, and contention at shared bottlenecks. Link delay variation reflects effects such as route changes, queue buildup at intermediate nodes, and dynamic path selection. Importantly, we do not assume the presence of a deliberate or intelligent attacker manipulating these variables. Instead, we treat them as environmental factors that can naturally fluctuate across runs, potentially creating challenging or adverse conditions for the controller. This perspective allows us to evaluate how policies behave under difficult but plausible network dynamics that may arise in practice. To simplify the analysis and isolate effects, we focus on bandwidth control in the following experiments and do not vary link delay.




\subsection{Adversarial Agent Design}
\label{sec:env-design}

For environment manipulation, the adversarial agent follows the same high-level design (with the improved reward function) as the feature-level agent (Section~\ref{sec:feature-design}). The additional factors we need to considier are (1) whether and how to control the level of dynamics of the environment traces, and (2) how to establish a baseline reference for queuing delay (a constraint in the reward function, see Section~\ref{sec:feature-design}). 

\begin{figure}[t]
\centering
\begin{minipage}[t]{0.47\textwidth}
\includegraphics[width=0.99\textwidth]{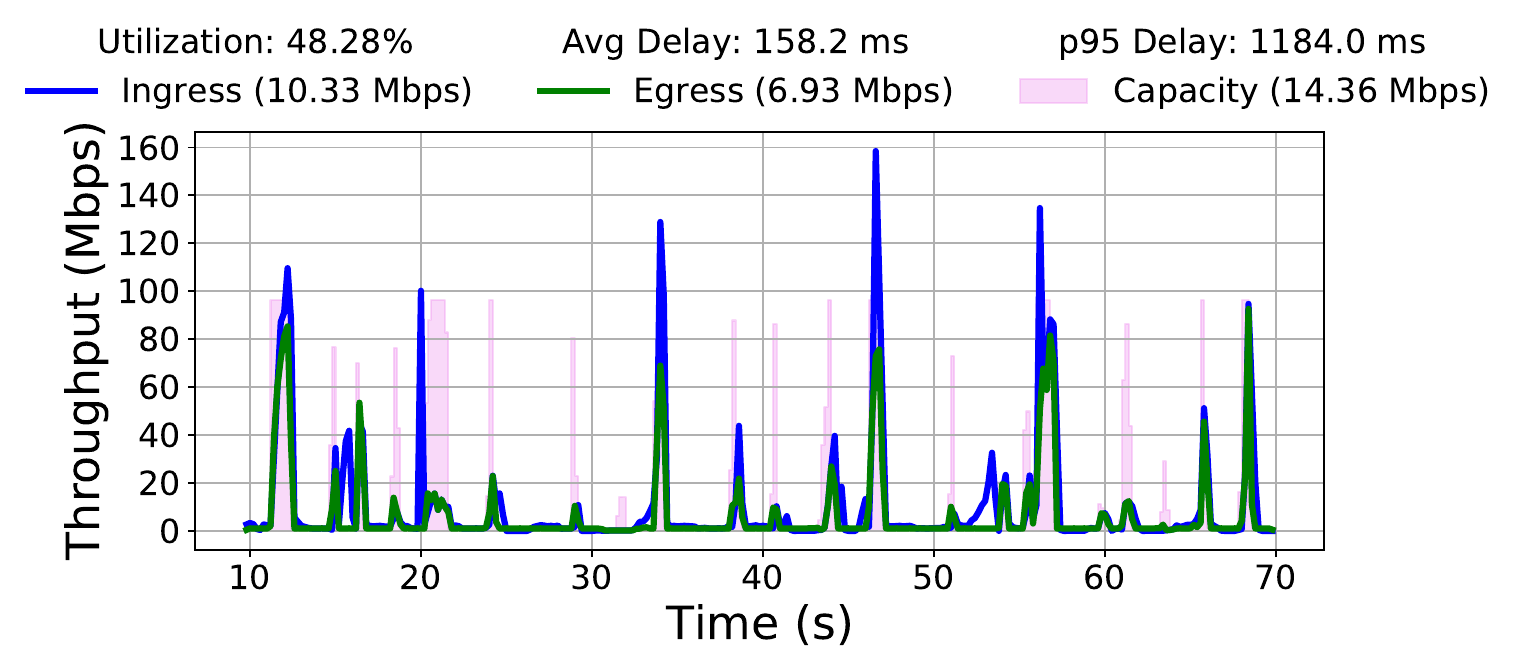}
\end{minipage}
\caption{
Example of an adversarially generated bandwidth trace targeting Orca under the unconstrained setting.
}
    \label{fig:env_orca_unconstrained}
    \vspace{-0.12in}
\end{figure}

\para{Unconstrained Adversarial Budget.}
We start with a {\em naive design} where the adversarial agent has no constraint on the bandwidth dynamics. In this setting, the adversarial agent is allowed to set an arbitrary bandwidth value at each time step within a predefined range (e.g., $[1, 96]$ Mbps, consistent with the range observed in the Canopy trace datasets). In this case, the adversarial agent learns to generate bandwidth traces that severely degrade the performance of congestion control policies. Figure~\ref{fig:env_orca_unconstrained} shows an example trace generated under this naive design, targeting Orca. While the traces successfully force a bad performance of the targeting CC (i.e., low bandwidth utilization and high queuing delay), the bandwidth traces also exhibit highly unrealistic dynamics, frequently oscillating between the minimum and maximum values with abrupt changes across consecutive time steps. For example, the bandwidth may jump from $1$ Mbps to $96$ Mbps within a single step and immediately drop back to $1$ Mbps in the next. Such traces will trivially force a bad performance, and limited insights can be drawn from the adversarial testing.

\begin{table*}[t]
    \centering
    \setlength{\tabcolsep}{6pt}
    \renewcommand{\arraystretch}{1.25}
    \begin{tabular}{lcccccc}
        \toprule
        \textbf{Tested Policy} & \textbf{Orca} & \textbf{Cubic} & \textbf{Vegas} & \textbf{Canopy} & \textbf{Illinois} & \textbf{LP} \\
        \midrule
        Baseline (random) & 95.58\% / 12.65ms & 94.50\% / 12.26ms & 65.16\% / 2.23ms & 89.69\% / 17.06ms & 98.69\% / 15.40ms & 85.58\% / 6.77ms \\
        \midrule
        \multicolumn{7}{l}{\textbf{Target Policy of Adv Agent}} \\
        \midrule
        Orca     & \underline{85.49\% / 61.54ms} & 95.37\% / 51.25ms & 66.68\% / 26.25ms & 87.67\% / 68.88ms & 95.84\% / 55.17ms & 90.52\% / 40.97ms \\
        Cubic    & 85.53\% / 72.42ms & \underline{86.79\% / 43.91ms} & 50.08\% / 17.23ms & 88.46\% / 79.48ms & 90.14\% / 46.07ms & 80.16\% / 33.24ms \\
        Vegas    & 90.88\% / 18.49ms & \textbf{75.68\% / 15.41ms} & \underline{\textbf{48.42\% / 7.48ms}}  & 84.78\% / 20.08ms & \textbf{80.25\% / 15.67ms} & 61.81\% / 13.14ms \\
        Canopy       & 90.11\% / 15.61ms & 76.70\% / 13.14ms & 49.96\% / 7.36ms  &  \underline{\textbf{82.72\% / 17.76ms}} & 83.27\% / 14.87ms & \textbf{55.95\% / 10.36ms} \\
        Illinois & \textbf{83.45\% / 22.41ms} & 81.04\% / 19.80ms & 55.41\% / 10.75ms & 85.37\% / 25.19ms  & \underline{83.60\% / 20.78ms} & 72.82\% / 15.48ms \\
        LP       & 93.36\% / 17.38ms & 87.06\% / 14.16ms & 54.27\% / 5.61ms  & 85.31\% / 20.46ms & 92.41\% / 17.39ms &  \underline{75.40\% / 10.26ms} \\
        \bottomrule
    \end{tabular}
    \caption{Environment-level adversarial testing results under bandwidth-trace manipulation. Budget=48 Mbps, bw\_range=[1,96]. The \underline{underlined} cells refer to the setting where the adversarial agent was trained and tested on the same CC. \textbf{Bold} cells denote the strongest attacks that degrade the bandwidth utilization of a tested CC in each column.}
    \label{tab:env_bw}
\end{table*}

\para{Constrained Adversarial Budget.}
Motivated by the above observations, we impose explicit constraints on the adversary's ability to manipulate bandwidth traces. More specifically, we introduce a budget-based constraint that enforces temporal smoothness in the generated traces. Specifically, let $b_t$ denote the bandwidth at time step $t$. We define the \emph{average absolute slope} over a sliding window of size $k$ (including the current step) as:
\begin{equation}
S_t = \frac{1}{k} \sum_{i=t-k+1}^{t} \left| b_i - b_{i-1} \right|.
\end{equation}
This metric captures the average magnitude of bandwidth changes between consecutive time steps within the window.

We then enforce a smoothness budget $\delta$ such that the average absolute slope $S_t \leq \delta$ at all time steps. During trace generation, each new bandwidth value $b_t$ is selected to ensure that the resulting sequence satisfies this constraint. This design prevents unrealistic high-frequency oscillations while still allowing the adversarial agent to generate sufficiently diverse and challenging network conditions. This allows protocol developers to generate traces that better reflect plausible real-world dynamics and provide more meaningful insights into the robustness of congestion control policies.


\para{Random Baseline.}
Unlike feature-level manipulation, we don't have a clear baseline to compare against, since the agent directly manipulates the bandwidth traces. Here, we need to establish a baseline for queuing delay to act as the constraint for the reward function. 
Our idea is to construct a {\em random baseline} by generating bandwidth traces that follow the same constraints as the adversarial setting, {\em but without any optimization}. At each time step, the bandwidth is randomly sampled within the same range (e.g., $[1, 96]$ Mbps). To ensure fairness and realism, these traces are subject to the same smoothness constraint, i.e., the average absolute slope $S_t \leq \delta$ is enforced during generation. This ensures that the random traces exhibit comparable temporal dynamics.

We generate 10 such random traces and evaluate each congestion controller on these traces, repeating each run three times to account for variability. The reported random baseline is obtained by averaging across all runs and traces. This baseline reflects typical performance under constrained but non-adversarial network variations, and provides a reference point for measure the effect of adversarial optimization.

\subsection{Evaluation Setup}

We first train a dedicated adversarial agent to manipulate each of the target CCs. During training, the agent learns to generate bandwidth sequences that degrade the performance of a specific CC under the reward design described earlier.

For evaluation, we first deploy the trained adversarial agent online. More specifically, while the target CC is running and interacting with the environment, the corresponding adversarial agent observes the current environment feedback at each time step and generates the bandwidth value for the next step according to its learned policy. The generated bandwidth values are bounded within a predefined range of 1~Mbps to 96~Mbps, which reflects the typical operating conditions observed in the training datasets. This constraint ensures that adversarial traces remain within a realistic regime while still allowing sufficient flexibility to construct challenging conditions.

We repeat this process to obtain multiple adversarially generated bandwidth traces. Among the traces of each CC, we select the one that results in the worst performance for the target CC, subject to the delay constraint. This trace is then used as a fixed test environment and applied to all other CCs to evaluate {\em transferability}. We report the average performance of each policy over three runs on the selected trace to ensure stability of the results.

As a baseline, we generate random bandwidth traces that satisfy the same smoothness and range constraints as the adversarial traces. Each policy is evaluated on these random traces, and we report the average performance across runs. This baseline reflects typical performance under non-adversarial but dynamic network conditions and serves as a reference point for measuring adversarial impact.

All other environment parameters remain consistent with the settings described earlier~\ref{sec:feature_level}.

\begin{figure}[t]
\centering
\begin{minipage}[t]{0.4\textwidth}
\includegraphics[width=0.99\textwidth]{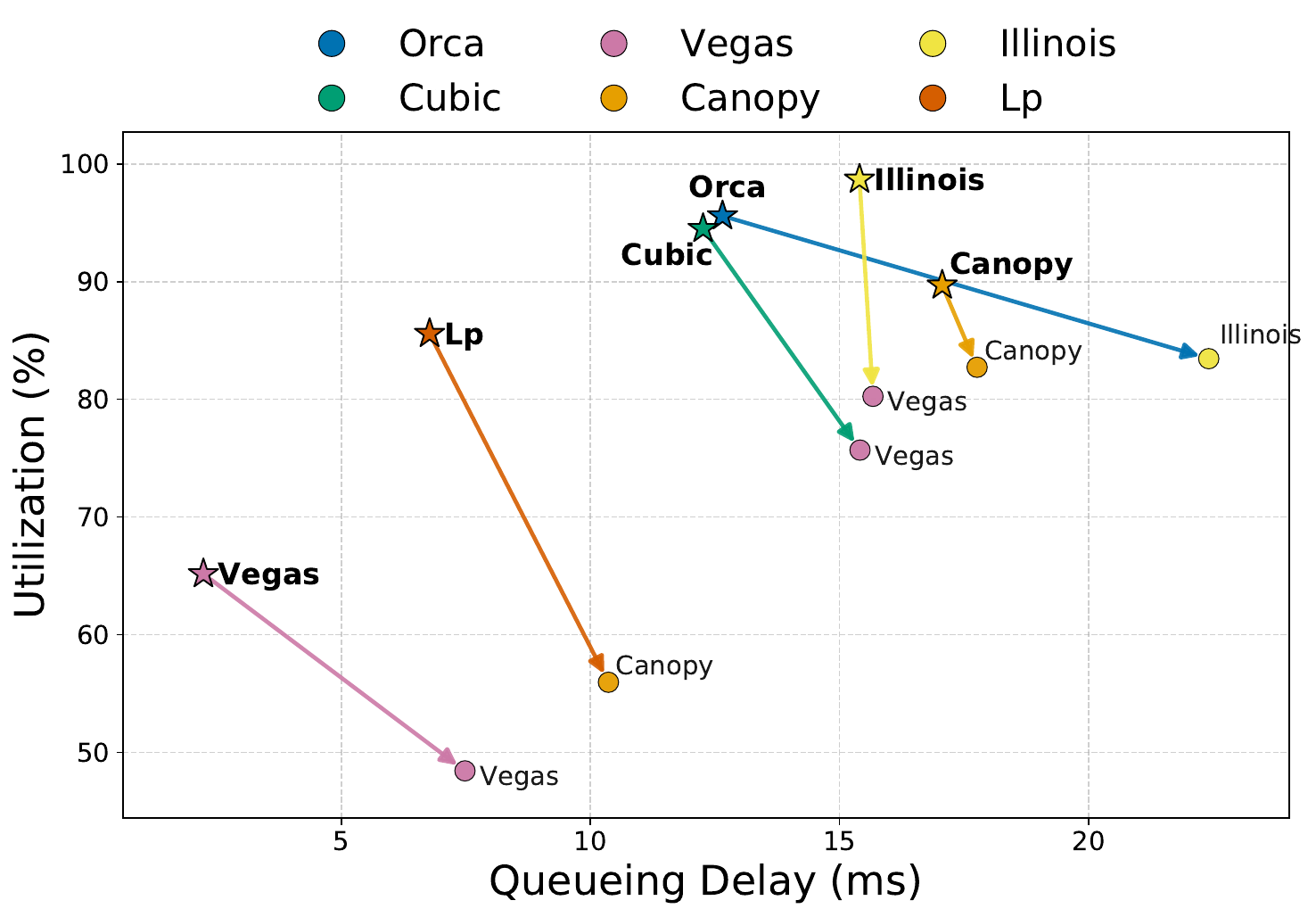}
\end{minipage}
\caption{
Environment-level adversarial testing results. We only show adversarial traces that lead to the strongest utilization degradation. The star represents performance on the random baseline, and the circle represents the performance on the strongest adversarial traces. 
}
    \label{fig:env_space}
    \vspace{-0.12in}
\end{figure}
\subsection{Performance and Findings}

Table~\ref{tab:env_bw} and Figure~\ref{fig:env_space} summarize the environment-level adversarial testing results under bandwidth-trace manipulation.

First, in Table~\ref{tab:env_bw}, the diagonal entries (underlined) represent the setting where the adversarial agent is trained against a target CC and evaluated on the same CC. We observe that all adversaries successfully achieve their objective.  In each case, the generated bandwidth traces induce queuing delay above the predefined threshold while simultaneously degrading utilization compared to the random baseline. This confirms that the proposed reward design is effective in guiding the adversarial agent to identify challenging network conditions that expose performance weaknesses.

Interestingly, we find that targeted adversarial traces are not always the most effective in degrading the utilization of a given CC. As shown in Table~\ref{tab:env_bw}, the off-diagonal entries represent transferred attacks where a trace generated against one target CC is evaluated on other CCs. 
The worst-case utilization for each tested policy is highlighted in bold. We find that in several cases, these transferred traces result in even lower utilization than the directly optimized ones, while still maintaining queuing delay above the threshold. We also observe that these transferred cases often have lower queuing delay, which likely helps to lower the bandwidth utilization even more. Overall, the result suggests that adversarial conditions discovered for one policy can generalize and may expose greater vulnerabilities in others. 


Finally, we quantify the maximum utilization degradation under the delay constraint for each policy. Learning-based policies (Orca and Canopy) exhibit relatively smaller degradation, around $12\%$ and $7\%$, respectively. In contrast, non-learning-based policies such as Cubic, TCP Vegas, TCP Illinois, and TCP LP experience significantly larger degradation, about $19\%$, $17\%$, $18\%$, and $30\%$, respectively. Overall, these results indicate that, under the considered adversarial setting, learning-based congestion controllers demonstrate stronger robustness compared to traditional non-learning-based ones.

\begin{tcolorbox}[width=\linewidth, colback=white!95!black, boxrule=0.5pt, left=2pt,right=2pt,top=1pt,bottom=1pt]
\stepcounter{remark}
{\bf Remark \arabic{remark}:}
{
Under environment-level bandwidth manipulation, learning-based CCs are also more robust than non-learning-based CCs, exhibiting smaller worst-case degradation. Also, we observe that {\em transferred} adversarial traces sometimes have a stronger impact than the {\em targeted} traces on degrading the performance of the given CCs.
}
\end{tcolorbox}


\subsection{Why Learning-based Policies Are More Robust}
\label{sec:env_reason}

The results in the previous subsection show that learning-based policies are consistently harder to degrade under our adversarial testing framework, exhibiting stronger robustness than non-learning-based designs. A natural question is \emph{why}.

\begin{table}[t]
    \centering
    \begin{tabular}{lcc}
        \toprule
        \textbf{Policy} & \textbf{cwnd Smoothiness} & \textbf{cwnd Smoothines (log)} \\
        \midrule
        Orca  & 6544.59  & 7.13  \\
         Canopy & 7149.33 & 7.34 \\
        \midrule
        Cubic  & 180.13 & 2.15 \\
        Vegas & 157.85 & 2.28 \\
        Illinois & 951.42 & 4.46 \\
        LP & 900.11 & 4.33 \\
        \bottomrule
    \end{tabular}
    \caption{Average cwnd Smoothiness different policies running on adversarial generated traces (Canopy targeted)}
    \label{tab:cwnd_smoothiness}
\end{table}

\para{Smoothness in Action Space?}
One hypothesis is that non-learning-based, rule-driven policies rely on fixed thresholds and pre-defined control rules, which may lead to relatively discrete congestion window (cwnd) adjustments. In contrast, learning-based policies operate over a more flexible and continuous action space, potentially enabling smoother control behavior. To verify this hypothesis, we record the cwnd evolution of all policies when running on the same adversarially generated trace. For this analysis, we focus on the adversarial trace targeting Canopy since all policies under this trace experience noticeable utilization degradation. To quantify cwnd dynamics, we first adopt the average absolute slope metric defined earlier. Additionally, since some policies may produce extremely large cwnd values, we introduce a log-scaled version to mitigate the impact of magnitude:
\begin{equation}
S_t^{\log} = \frac{1}{k} \sum_{i=t-k+1}^{t} 
\left| \frac{\log(b_i) - \log(b_{i-1})}{t_i - t_{i-1}} \right|.
\end{equation}

The results are summarized in Table~\ref{tab:cwnd_smoothiness}. Surprisingly, learning-based policies exhibit \emph{significantly larger} cwnd variation. For example, Orca and Canopy achieve values of 6544.59 and 7149.33, respectively, while all non-learning-based policies remain below 1000. A similar trend is observed under the log-scaled metric. These results {\em do not support} the initial hypothesis: learning-based policies do not have smoother changes in the congestion window; instead, they exhibit much more aggressive and dynamic cwnd adjustments.

\begin{figure}[t]
\centering
\begin{minipage}[t]{0.45\textwidth}
\includegraphics[width=0.99\textwidth]{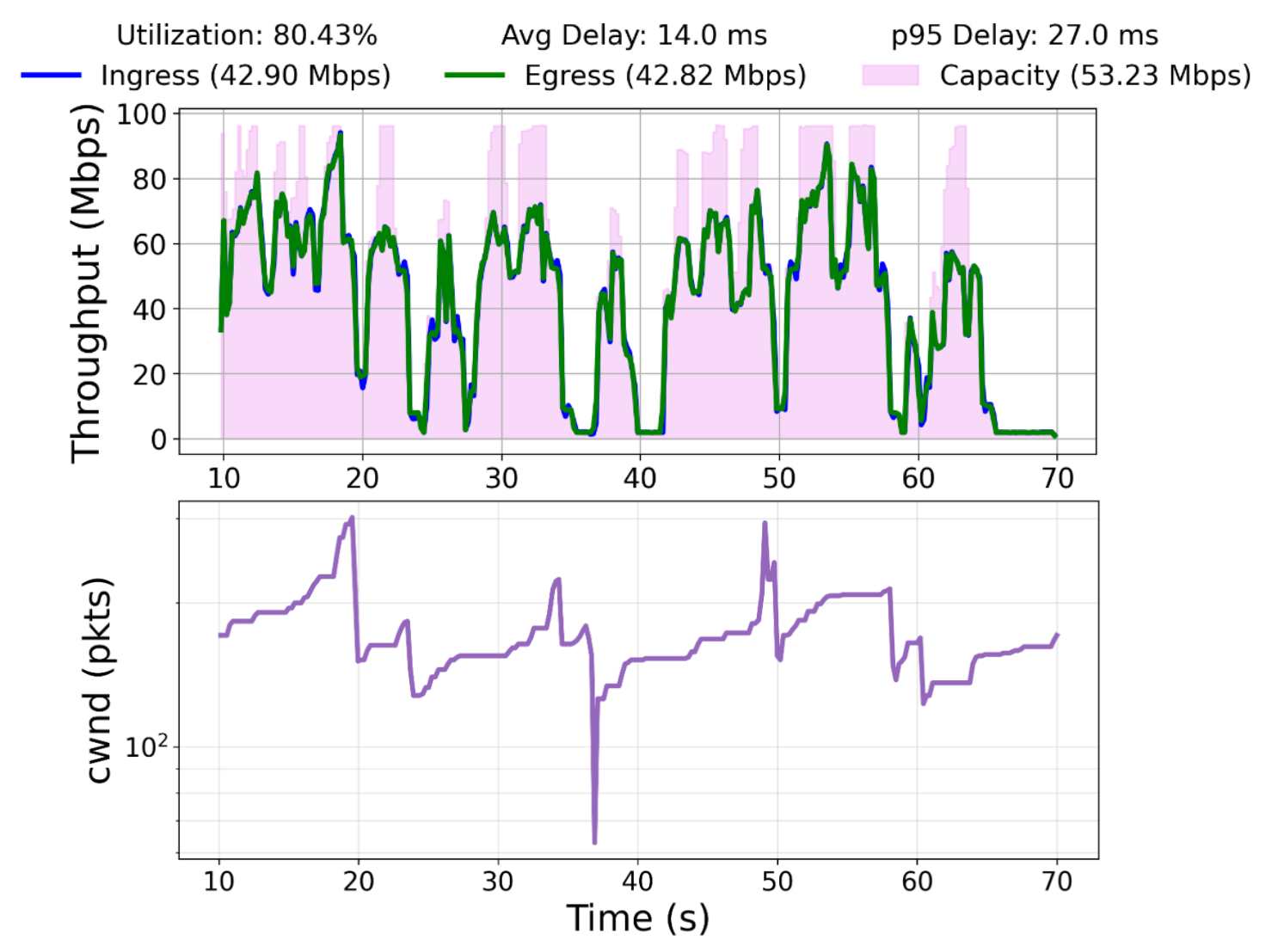}
\end{minipage}
\caption{
A comparison example between Cubic's egress and ingress curves and cwnd curve on adversarial-generated bandwidth traces (Canopy targeted). 
}
    \label{fig:cwnd_cubic}
    \vspace{-0.12in}
\end{figure}

\begin{figure}[t]
\centering
\begin{minipage}[t]{0.45\textwidth}
\includegraphics[width=0.99\textwidth]{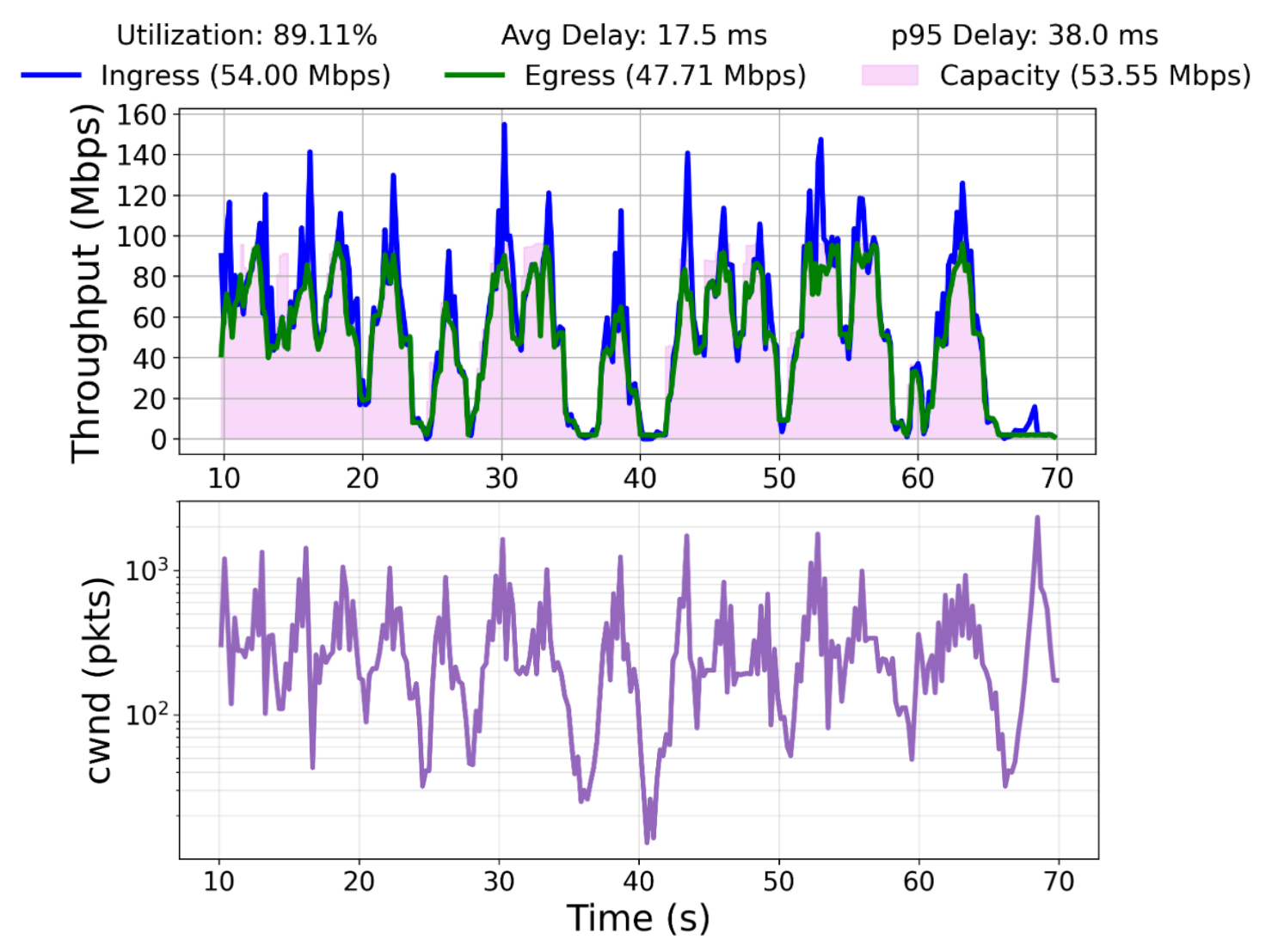}
\end{minipage}
\caption{A comparison example between Orca's egress and ingress curves and cwnd curve on adversarial-generated bandwidth traces (Canopy targeted). }
    \label{fig:cwnd_orca}
    \vspace{-0.12in}
\end{figure}

\para{Responsiveness of CWND Adjustments?}
To further understand the differences between learning-based and non-learning-based  CCs, we further examine the cwnd adjustments of these CCs under the same adversarial traces. Figure~\ref{fig:cwnd_cubic} and Figure~\ref{fig:cwnd_orca} show an example. Each figure compares the ingress and egress throughput curves (upper panel) with the cwnd evolution (lower panel). For Cubic (Figure~\ref{fig:cwnd_cubic}), the cwnd can only loosely follow the bandwidth dynamics, showing limited responsiveness to rapid changes. In contrast, Orca (Figure~\ref{fig:cwnd_orca}) exhibits a much stronger alignment: when the ingress/egress curves need a peak to match the burst in available bandwidth, the cwnd will correspondingly increase, and vice versa. 

These observations indicate that the advantage of learning-based policies does not come from smoother control, but from their ability to perform timely cwnd adjustments (even if some adjustments are drastic). Their more flexible action space enables rapid adaptation to dynamic conditions, allowing them to better track bandwidth variations under adversarial traces, whereas rule-based designs remain constrained by their predefined control mechanisms.

\begin{tcolorbox}[width=\linewidth, colback=white!95!black, boxrule=0.5pt, left=2pt,right=2pt,top=1pt,bottom=1pt]
\stepcounter{remark}
{\bf Remark \arabic{remark}:}
{
Based on our cwnd analysis, learning-based CCs do not rely on smoother cwnd adjustments; instead, they exhibit more aggressive (higher-variation) cwnd dynamics. This enables more responsive adaptation to dynamic network conditions, leading to improved robustness.
}
\end{tcolorbox}

\section{A Case Study of TCP LP}
\label{sec:case_study}

The previous section shows that adversarially generated bandwidth traces can significantly degrade utilization across different congestion control policies. Beyond quantitative evaluation, an important question is what insights can be extracted from these traces. In this section, we perform a case study on TCP LP to understand how specific algorithmic design choices interact with adversarially induced network dynamics.

\subsection{Overview of TCP LP}

TCP LP is a delay-based congestion control algorithm built on top of TCP Reno, with an additional early congestion detection mechanism. Importantly, TCP LP is designed to operate as a low-priority background flow: its goal is to utilize only excess bandwidth without interfering with coexisting standard TCP connections, and to yield quickly when congestion is detected. 

To understand TCP LP, we first describe how TCP Reno works. TCP Reno operates in three main phases: slow start, congestion avoidance, and fast recovery. When the congestion window ($cwnd$) is below the slow start threshold ($ssthresh$), it increases exponentially, adding one MSS per ACK, resulting in exponential growth per RTT. Once $cwnd \geq ssthresh$, TCP Reno enters congestion avoidance, where $cwnd$ increases linearly according to:
\begin{equation}
cwnd \leftarrow cwnd + \frac{\text{MSS}^2}{cwnd}.
\end{equation}
When three duplicate ACKs are received, the sender enters the fast recovery state. At this point, the congestion window is reduced by setting $ssthresh = cwnd/2$ and adjusting $cwnd$ accordingly. The sender retransmits the lost packet and continues transmitting new data upon receiving duplicate ACKs to keep the pipeline full. Once the retransmitted packet is acknowledged, $cwnd$ is set to $ssthresh$, and the sender exits fast recovery and resumes congestion avoidance.

TCP LP inherits all TCP Reno states but introduces an additional layer of control based on delay signals to proactively detect congestion. In addition to standard loss-based mechanisms, TCP LP continuously monitors one-way delay (OWD), which reflects the time packets spend in the network, and its smoothed version (SOWD) to reduce noise and transient fluctuations. By maintaining estimates of the minimum and maximum observed delay, TCP LP establishes a baseline for propagation delay and tracks deviations caused by queue buildup. The key intuition is that increasing delay indicates growing queues in the network, which often precedes packet loss; thus, TCP LP attempts to anticipate congestion before losses occur. When the smoothed delay exceeds a predefined threshold, TCP LP interprets this as early congestion and reacts proactively by reducing the congestion window, either by halving it or, in more severe cases, resetting it to a very small value (e.g., $cwnd=1$). This design allows TCP LP to yield bandwidth earlier than traditional loss-based algorithms to minimize queuing delay, but also makes it more sensitive to delay fluctuations, which can lead to conservative bandwidth usage under dynamic or adversarial network conditions.

\subsection{Behavior under Adversarial Traces}

\begin{figure}[t]
\centering
\begin{minipage}[t]{0.45\textwidth}
\includegraphics[width=0.99\textwidth]{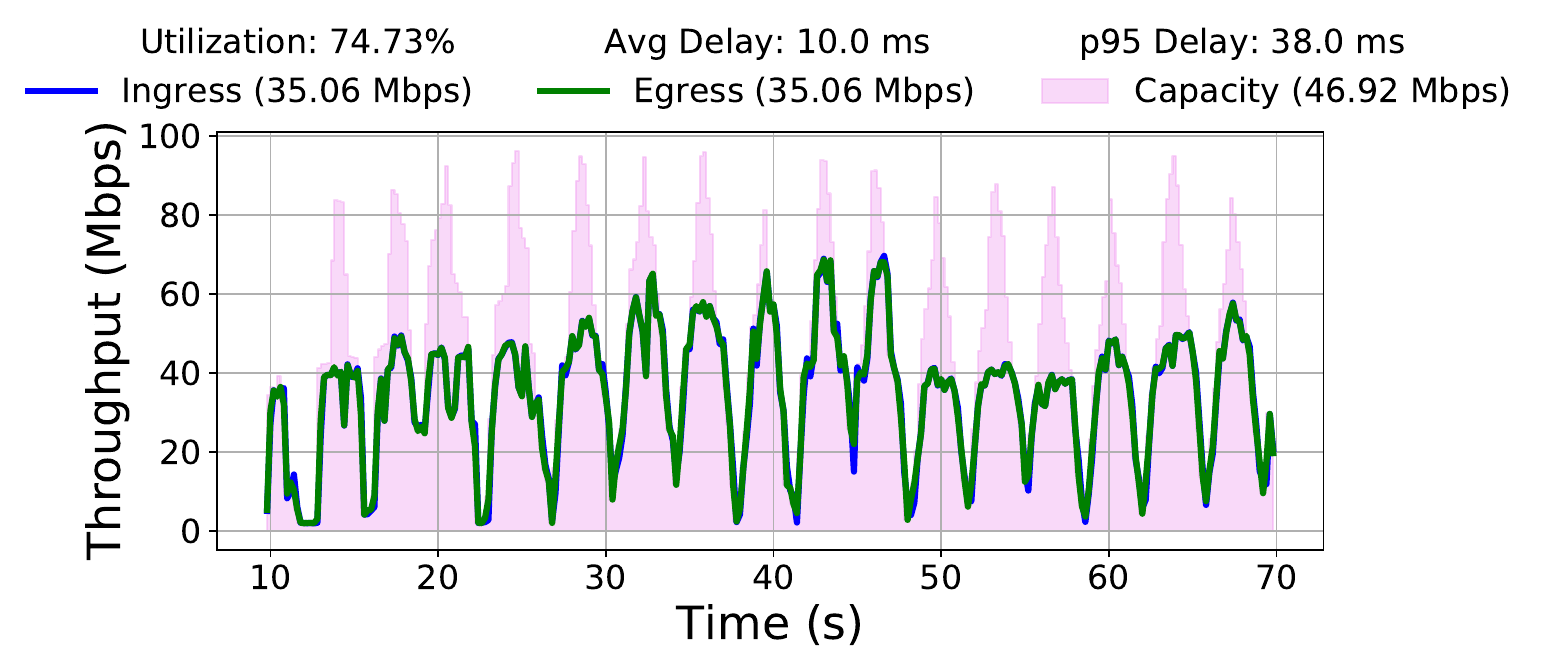}
\end{minipage}
\caption{
Visualization of one testing case of TCP LP.
}
    \label{fig:lp_example}
    \vspace{-0.12in}
\end{figure}

Using the adversarially generated bandwidth traces (Figure~\ref{fig:lp_example}), we observe a recurring burst-like pattern in available capacity, where bandwidth gradually increases to a peak and is then followed by a sustained decrease to a lower level, repeating over time. This structured variation has a distinct impact on TCP LP’s behavior.

During the initial several bursts, TCP LP operates in the congestion avoidance phase. No packet loss is observed, and the delay signal (OWD/SOWD) does not exceed the early congestion threshold. As a result, the congestion window increases linearly following TCP Reno’s congestion avoidance rule. However, due to the rise-and-decline pattern of the bandwidth, the sending rate does not fully adapt to the capacity reduction phase following each peak. This mismatch causes packets to accumulate in the queue, leading to a gradual buildup of queuing delay across successive cycles.

\begin{figure}[t]
\centering
\begin{minipage}[t]{0.45\textwidth}
\includegraphics[width=0.99\textwidth]{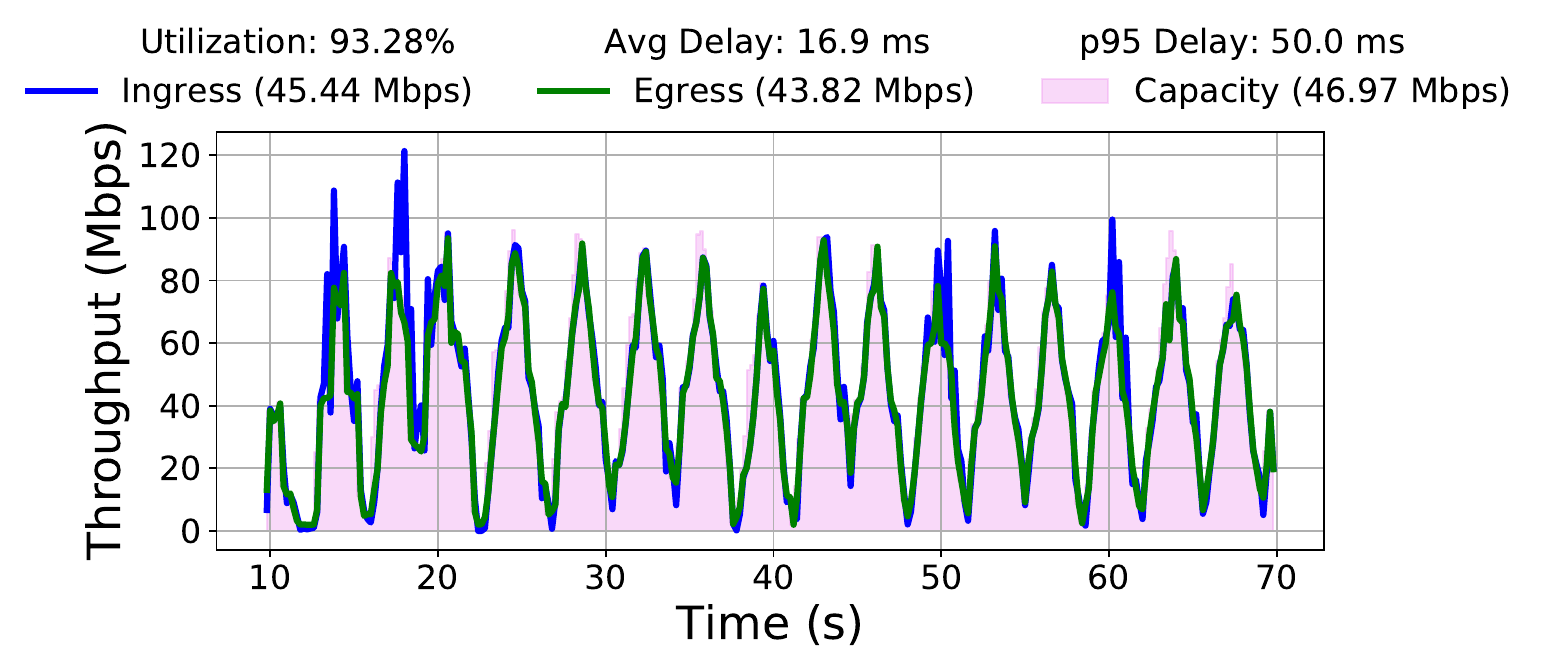}
\end{minipage}
\caption{
Replaying the TCP LP adversarial trace on Orca.
}
    \label{fig:lp_on_orca}
    \vspace{-0.12in}
\end{figure}

Although each individual triangle does not immediately trigger TCP LP’s early congestion detection, the accumulated effect across multiple cycles causes both OWD and its smoothed version (SOWD) to increase progressively. Eventually, at a later triangle, the smoothed delay exceeds the predefined threshold. At this point, TCP LP interprets the condition as early congestion and triggers a proactive response, sharply reducing the congestion window (e.g., halving it), even though packet loss has not yet occurred.

This behavior highlights a key characteristic of TCP LP under such adversarial traces: the algorithm does not react to instantaneous bandwidth decreases within a single cycle, but instead responds to the cumulative delay buildup over time. As a result, TCP LP may delay its reaction until queuing effects become pronounced, leading to abrupt window reductions that are not directly aligned with the instantaneous available bandwidth.

\subsection{Insights}

This behavior reveals two key insights into TCP LP’s design under such dynamics. First, TCP LP relies on delay as a signal of congestion, where increasing delay is interpreted as evidence of queue buildup in the network. However, this signal reflects the accumulation of packets over time rather than the instantaneous availability of bandwidth. In the adversarial traces, when bandwidth increases, the queue does not immediately drain, and the measured delay remains elevated due to packets buffered during earlier mismatches between sending rate and capacity. As a result, TCP LP continues to interpret that the network is congested even during periods when additional bandwidth has become available. Instead of probing this newly available capacity, the protocol reacts based on the elevated delay signal, which effectively captures past conditions rather than the current state of the network. This leads to situations where TCP LP refrains from increasing its sending rate, even though the underlying capacity would permit further utilization. 

Second, once the delay signal exceeds the early congestion threshold, TCP LP applies abrupt reductions to the congestion window. While this behavior is consistent with its design goal of yielding quickly to competing traffic, it can be overly conservative under dynamic conditions such as the burst-like patterns observed here. In these environments, capacity fluctuates frequently, and short-lived increases in bandwidth require timely and gradual adaptation to be effectively utilized. However, TCP LP’s sharp backoff interrupts this process. After that, TCP LP must rebuild its congestion window through slow linear growth---before it can reach the available bandwidth, another cycle of delay buildup triggers early congestion again, causing it to back off once more. 

As a comparison, we replay the same adversarial trace on Orca (a learning-based CC), as shown in Figure~\ref{fig:lp_on_orca}. Under the same bandwidth trace, Orca achieves substantially higher utilization (93.28\%) by adapting more flexibly to the burst-like bandwidth dynamics. 
Note that this is not to criticize the design of TCP LP. Instead, the results can help protocol developers to validate their design and help to inform practitioners regarding which scenarios a given protocol may not work well. 


\begin{tcolorbox}[width=\linewidth, colback=white!95!black, boxrule=0.5pt, left=2pt,right=2pt,top=1pt,bottom=1pt]
\stepcounter{remark}
{\bf Remark \arabic{remark}:}
{
Adversarial traces can be used to reveal failure modes of a congestion controller and analyze how the design choices affect its performance under challenging network conditions. 
}
\end{tcolorbox}

\section{Adversarial Training}
\label{sec:adv_train}

Our analysis so far shows that adversarially generated traces can identify failure modes of non-learning-based CCs by revealing subtle mismatches between control logic and network dynamics. For learning-based policies, such traces provide another opportunity: by incorporating the challenging traces into protocol training, we expect to further improve the learning-based CCs. 

For this experiment, we focus on Orac since Orac does not have any robustness-enhancing designs. Orca, by default, is trained on samples collected from relatively stable bandwidth environments, which explains why its policy suffers under adverse dynamics. In comparison, Canopy already has a built-in robustness module (e.g., formally verified learning with robustness as a desired property). We will explore whether we can improve Orac's robustness with an adversarial training strategy, to match or exceed that of Canopy.   


\subsection{Experiment Setup}

To evaluate adversarial training, we modify Orca's original training pipeline to explicitly control the exposure of the agent to adversarial traces. Instead of training from scratch, we initialize from the pre-trained model and continue training with an augmented trace pool.

\para{Trace Integration.}
In the original Orca training setup, multiple agents are typically trained in parallel, each interacting with a fixed bandwidth trace. To better control the training distribution, we abstract this process into a unified sampling framework. Specifically, rather than binding each agent to a fixed trace, we maintain a shared trace pool and allow the training process to dynamically select traces. During training, the agent interacts with one selected trace at a time for a fixed number of steps to collect experience, after which a new trace may be selected. We incorporate adversarial traces into this pool alongside the original benign traces, enabling the agent to experience both types of network dynamics within a single training process. Importantly, aside from the composition of the trace pool, all other components of the original training framework remain unchanged. In particular, the reward function, model architecture, and optimization procedure are identical to those used in the original Orca training setups. 

\para{Mixing Ratio.}
Within this unified framework, the mixing ratio is defined as the probability of selecting an adversarial trace when sampling from the trace pool. At each trace switch, an adversarial trace is chosen with probability $p$, while a benign trace is selected with probability $1-p$. This abstraction allows us to directly control how frequently the agent is exposed to adversarial conditions, independent of the number of traces or parallel agents. The mixing ratio $p$ is treated as a hyperparameter, and a detailed study of its impact is provided in Appendix~\ref{sec:mixratio}.

\subsection{Performance and Findings}


We evaluate the effectiveness of adversarial training under the setup described above. Based on an analysis of the mixing ratio (Appendix~\ref{sec:mixratio}), we set $p = 0.2$, which provides a good balance between robustness under adversarial conditions and performance on benign workloads. The results are shown in Table~\ref{tab:adv_retrain}.

Overall, adversarial training does not significantly degrade performance on benign environments. On both the simulated and real-world evaluation datasets~\cite{yang2024canopy}, the retrained policy achieves comparable utilization, with only a moderate increase in queuing delay. Similarly, performance on the random baseline remains largely unchanged, indicating that the policy preserves its general behavior under standard conditions.

In contrast, the improvement on the adversarial trace is substantial. The retrained policy achieves higher utilization (6\% improvement, from 85.49\% to 91.66\%) while reducing queuing delay compared to the original model. Also, the re-trained Orca even performs better than the Canopy, a robustness-enhanced protocol~\cite{yang2024canopy}.  
This suggests that exposure to adversarial dynamics during training enables the policy to better anticipate and adapt to structured bandwidth variations. 



\begin{tcolorbox}[width=\linewidth, colback=white!95!black, boxrule=0.5pt, left=2pt,right=2pt,top=1pt,bottom=1pt]
\stepcounter{remark}
{\bf Remark \arabic{remark}:}
{
Adversarially generated traces expose hard cases that are underrepresented in standard training data. By incorporating these traces during training, learning-based CCs can achieve improved robustness under challenging network conditions without sacrificing performance on normal workloads.
}
\end{tcolorbox}

\begin{table}[t]
    \centering
    \begin{tabular}{lllcc}
        \toprule
        \textbf{Model} &\textbf{Model} & \textbf{Settings} & \textbf{Util. ($\uparrow$)} & \textbf{Delay ($\downarrow$)} \\
        \midrule
        \multirow{8}{*}{Orca} & \multirow{2}{*}{Random Baseline} 
               & No Retrain  & 95.58 & 12.65 \\
           &   & Adv Train & 94.65 & 12.86 \\
        \cmidrule{2-5}
        & \multirow{2}{*}{Synthetic Traces~\cite{yang2024canopy}} 
               & No Retrain  & 94.11 & 14.18 \\
           &   & Adv Train & 95.12 & 21.03 \\
         \cmidrule{2-5}
        & \multirow{2}{*}{Real-world Traces~\cite{yang2024canopy}} 
               & No Retrain  & 83.21 & 29.41 \\
           &   & Adv Train & 83.89 & 30.09 \\
         \cmidrule{2-5}
        & \multirow{2}{*}{Adversarial Trace} 
               & No Retrain  & 85.49 & 61.54 \\
           &  & Adv Train & 91.66 & 57.83 \\
        \midrule
        Canopy & Adversarial Trace & No Retrain & 86.67 & 68.99 \\
        \bottomrule
    \end{tabular}
    \caption{Before and after adversarial training for Orca. The metrics are now displayed from the defenders' perspective (e.g., higher is better for utility).  
    }
    \label{tab:adv_retrain}
\end{table}

\section{Discussion and Conclusion}

We introduce \sysname{}, an adversarial testing framework for evaluating and comparing learning- and non-learning-based congestion controllers (CCs). At the feature level, we examine the sensitivity of CCs to perturbations on shared input signals. At the environment level, we generate adversarial traces to evaluate how controllers behave under challenging network conditions. Through our experiments, we observe that learning-based congestion controllers exhibit greater robustness than traditional human-designed controllers. Our analysis further shows that this robustness does not stem from smoother control behavior, but from the ability to adapt more responsively to dynamic network conditions. Finally, we show adversarial training can help further improve the robustness of learning-based CCs. 

Future work may explore broader use cases of CCLab. In this paper, we demonstrate adversarial testing that targets bandwidth utilization degradation. However, other attack targets (e.g., against fairness, queuing delay) can be implemented into CCLab by simply changing the {\em reward function} of the adversarial agent. In addition, future work may use CCLab to perform adversarial tests on other input signals (e.g., acknowledgment statistics) to cover other categories of CCs (e.g., loss-based CCs). Finally, future testing may consider manipulating other network environment conditions, such as link delay. Our evaluation is focused on bandwidth manipulation as it has a direct impact on the congestion controller. Intuitively, link delay should not have the same impact, and future work is needed to explicitly compare their impacts on CCs' behaviors. 

\newpage
\section*{Ethics Considerations}
\label{sec:ethics}

This work evaluates congestion control algorithms in a fully controlled and emulated environment using Mahimahi, based on publicly available network traces. No experiments are conducted on live systems, and no real users or production traffic are affected. 
The adversarial testing framework is designed for evaluation and diagnostic purposes, rather than enabling real-world attacks. Its purpose is to identify failure modes and improve robustness.
That being said, we acknowledge potential concerns regarding misuse of the framework. We believe the benefit of this work (enabling robustness testing for protocol developers) outweighs the potential risk. We will share our code with the broad community to facilitate result reproduction and protocol testing, and support future research. However, to prevent misuse, we will verify the identity of the code requester (e.g., based on their institutional email) before sharing the code.  
Finally, this work does not involve the collection or analysis of personally identifiable information. The datasets used consist of network-level traces without user identities or sensitive content. Overall, our approach follows established ethical guidelines for ICT research, including minimizing harm and ensuring responsible use of experimental methods~\cite{kenneally2012menlo}.







\bibliographystyle{IEEEtran}
\bibliography{sample}

\appendix

\section{Appendix}
\label{sec:appendix}


\subsection{Feature-Level Manipulation: Real-world Traces}
\label{sec:feature_level_real}
To validate the generality of our findings beyond Canopy simulated traces, we repeat the feature-level adversarial experiments on Canopy real-world traces. We also apply the redesigned reward that constrains queuing delay and isolates utilization degradation. The corresponding results are shown in Tables~\ref{tab:feature_minrtt_005_real_newreward} and~\ref{tab:feature_minrtt_05_real_newreward}. Under this setting, queuing delay in adversarial cases remains comparable to the clean baseline, enabling a more direct comparison of robustness.

Under small perturbations (±5\%), all policies exhibit minimal utilization changes, indicating robustness to mild noise. As perturbation strength increases to ±50\%, clearer differences emerge. Learning-based policies (Orca and Canopy) show relatively moderate degradation, while non-learning-based policies exhibit more pronounced sensitivity. In particular, TCP Vegas experiences the largest drop in utilization (11\%), and BBR also shows noticeable degradation (7\%), whereas learning-based designs maintain more stable performance.

Overall, the results on real-world traces are consistent with those observed on synthetic datasets. When controlling for the utilization–delay trade-off, learning-based CCAs continue to demonstrate stronger robustness to feature-level perturbations, while non-learning-based policies—especially those relying directly on RTT signals—are more susceptible to adversarial manipulation.

\subsection{Feature-Level Manipulation: Noise Setting}
\label{sec:feature_level_noise}

In addition to adversarial perturbations, we consider a noise-based setting where the target feature (minimum RTT) is randomly perturbed within the same bounded range. Table~\ref{tab:noise_only} summarizes the results.

Unlike adversarial perturbations, random noise does not provide control over the utilization–delay trade-off. As a result, the system may shift toward different operating regimes, including (i) lower utilization with lower delay (more conservative behavior), (ii) higher utilization with higher delay (more aggressive behavior), or (iii) simultaneous degradation in both metrics. For example, under noise, BBR exhibits a noticeable drop in utilization accompanied by reduced queuing delay, reflecting a shift toward a more conservative operating point rather than a clear degradation. Overall, this lack of directional control leads to scattered and less interpretable outcomes.

\begin{table}[t]
    \centering
    \begin{tabular}{llccc}
        \toprule
        \textbf{Model} & \textbf{Settings} & \textbf{Util. ($\downarrow$)} & \textbf{Delay ($\uparrow$)} & \textbf{P95 delay ($\uparrow$)} \\
        \midrule
          Orca   & Clean    & 83.21 & 29.41 & 81.67 \\
           
             & Adversarial & 83.17 & 31.63 & 87.21\\
        \midrule
          Canopy   & Clean    & 91.22 & 39.84 & 112.14    \\
           
             & Adversarial  & 92.58 & 45.79 & 133.33\\
        \midrule
        BBR    & Clean  & 90.09 & 29.10 & 82.19\\
            
             & Adversarial & 85.45 & 29.16 & 84.14\\
        \midrule
        Vegas     & Clean & 79.98 & 12.53 & 34.08\\
           
             & Adversarial  & 77.88 & 12.44 & 39.17\\
        \bottomrule
    \end{tabular}
    \caption{Feature-space manipulation with real-world traces under minRTT perturbation ($[0.95, 1.05]$). In this experiment, the new reward is applied to maintain the queuing delay.
    }
    \label{tab:feature_minrtt_005_real_newreward}
\end{table}

\begin{table}[t]
    \centering
    \begin{tabular}{llccc}
        \toprule
        \textbf{Model} & \textbf{Settings} & \textbf{Util. ($\downarrow$)} & \textbf{Delay ($\uparrow$)} & \textbf{P95 delay ($\uparrow$)} \\
        \midrule
        Orca     & Clean   & 83.21 & 29.41 & 81.67 \\
         
             & Adversarial & 78.92 & 29.21 & 80.09 \\
        \midrule
         Canopy    & Clean   & 91.22 & 39.84 & 112.14   \\
           
             & Adversarial & 90.08 & 45.61 & 129.71\\
        \midrule
        BBR     & Clean  & 90.09 & 29.10 & 82.19\\
              
             & Adversarial & 83.13 & 30.69 & 90.50 \\
        \midrule
        Vegas     & Clean & 79.98 & 12.53 & 34.08\\
             
             & Adversarial & 68.53 & 13.29 & 38.75 \\
        \bottomrule
    \end{tabular}
    \caption{Feature-space manipulation with real-world traces under minRTT perturbation ($[0.5, 1.5]$). In this experiment, the new reward is applied to maintain the queuing delay.}
    \label{tab:feature_minrtt_05_real_newreward}
\end{table}

\begin{table}[t]
    \centering
    \setlength{\tabcolsep}{6pt}
    \renewcommand{\arraystretch}{1.2}
    \begin{tabular}{llccc}
        \toprule
        \textbf{Trace} & \textbf{Model} & \textbf{Util. ($\downarrow$)} & \textbf{Delay ($\uparrow$)} & \textbf{P95 delay ($\uparrow$)} \\
        \midrule
        \multirow{4}{*}{Simulated}
        & Orca   & 83.75 & 4.64  & 13.82 \\
        & Canopy & 93.35 & 19.10 & 45.54 \\
        & BBR    & 72.55 & 5.07  & 17.49 \\
        & Vegas  & 70.03 & 2.98  & 6.23 \\
        \midrule
        \multirow{4}{*}{Real-world}
        & Orca   & 84.51 & 29.39 & 81.83 \\
        & Canopy & 92.22 & 39.97 & 112.42 \\
        & BBR    & 81.03 & 16.31 & 49.37 \\
        & Vegas  & 75.27 & 11.68 & 31.67 \\
        \bottomrule
    \end{tabular}
    \caption{Noise-based perturbation. All results correspond to the noise setting, where perturbations are randomly sampled within the same range as the adversarial budget.}
    \label{tab:noise_only}
\end{table}

\begin{table}[t]
    \centering
    \begin{tabular}{ccccc}
        \toprule
        \textbf{Mixed Rate} & \multicolumn{2}{c}{\textbf{Random Baseline}} & \multicolumn{2}{c}{\textbf{Adv Train Performance}} \\
        \cmidrule(lr){2-3} \cmidrule(lr){4-5}
        & \textbf{Util. ($\downarrow$)} & \textbf{Delay ($\uparrow$)} & \textbf{Util. ($\downarrow$)} & \textbf{Delay ($\uparrow$)} \\
        \midrule
        0   & 95.58 & 12.65 & 85.49 & 61.54 \\
        0.1 & 82.32 & 7.73  & 80.11 & 33.50 \\
        0.2 & 90.82 & 8.70  & 91.66 & 57.83 \\
        0.5 & 69.59 & 6.57  & 88.74 & 41.00 \\
        0.8 & 33.72 & 2.97  & 90.35 & 39.03 \\
        1   & 34.15 & 4.00  & 57.23 & 18.66 \\
        \bottomrule
    \end{tabular}
    \caption{Adversarial training performance under different mixed rates.}
    \label{tab:mixed_rate}
\end{table}

\subsection{Adv. Training Mixing Ratio Influence}
\label{sec:mixratio}

To further understand the role of adversarial exposure during training, we vary the mixing ratio $p$ and evaluate the resulting policies across both benign and adversarial environments. The utilization and queuing delay trends results are summarized in Table~\ref{tab:mixed_rate}.

We observe that the retrained policy exhibits significant performance variance across different mixing ratios. In particular, when $p$ increases from 0.2 to 0.8, utilization and queuing delay under adversarial traces improve consistently, indicating that more frequent exposure to adversarial dynamics enables the policy to better adapt to such conditions. However, this gain comes at the cost of degraded performance on benign environments. As the mixing ratio increases, the random baseline performance deteriorates, with utilization dropping substantially, suggesting that excessive adversarial exposure biases the policy toward overly conservative or specialized behaviors that do not generalize well to standard conditions.

An interesting phenomenon appears at low mixing ratios. When $p = 0.1$, both utilization and queuing delay decrease compared to the baseline. While the reduction in delay may appear beneficial, it is accompanied by a noticeable drop in utilization, indicating under-utilization of available bandwidth. This behavior can be attributed to the original reward function used in Orca training, which inherently balances the trade-off between utilization and queuing delay. Since the reward formulation remains unchanged during adversarial retraining, the policy may respond to limited adversarial exposure by adopting a more conservative strategy, reducing sending rates to avoid potential congestion signals. As a result, both metrics decrease simultaneously, reflecting a shift along the utilization-delay trade-off rather than a clear performance improvement.

Overall, these results highlight that the mixing ratio plays a critical role in shaping the learning-based policy. Moderate values (e.g., $p = 0.2$) provide a better balance, improving robustness to adversarial conditions while maintaining reasonable performance on benign traces. In contrast, overly small or large mixing ratios can lead to suboptimal behaviors, either failing to capture adversarial dynamics or overfitting to them at the expense of generalization.

\end{document}